\documentclass[%
preprint,
nofootinbib,
 amsmath,amssymb,
prstab,
floatfix,
]{revtex4-1}
\usepackage{float}
\usepackage{graphicx}
\usepackage{comment}
\usepackage{natbib}

\newcommand{\mL}{\mathcal{L}}
\newcommand{\mN}{\mathcal{N}}
\newcommand{\mB}{\mathcal{B}}
\newcommand{\mP}{\mathcal{P}}
\newcommand{\mA}{\mathcal{A}}

\newcommand{\be}{{\boldsymbol e}}
\newcommand{\bF}{{\boldsymbol F}}

\newcommand{\bnabla}{{\boldsymbol \nabla}}
\newcommand{\br}{{\boldsymbol r}}

\newcommand{\bu}{{\boldsymbol u}}
\newcommand{\bU}{{\boldsymbol U}}

\newcommand{\ez}{{\boldsymbol e_z}}

\newcommand{\er}{{\boldsymbol e_r}}
\newcommand{\dt}{{\Delta t}}
\newcommand{\pd}{{\partial}}
\newcommand{\grad}{{\bnabla}}
\newcommand{\lap}{{\nabla^2}}

\newcommand{\curl}{{\bnabla\times}}
\newcommand{\divv}{{\bnabla \cdot}}
\newcommand{\nn}{{\nonumber}}
\renewcommand{\phi}{{\varphi}}
\newcommand{\Ra}{{\rm Ra}}
\newcommand{\Ek}{{\rm Ek}}
\newcommand{\Pra}{{\rm Pr}}
\newcommand{\Temp}{{T}}
\usepackage{subfig}
\begin{document}
\title{{\small \bf Cite as:
  J.C. Gonzalez Sembla, C. Rambert, F. Feudel, L.S. Tuckerman,
Mathematics (MDPI) {\bf 13}, 2113 (2025).}\\
Convection in a Rapidly Rotating Spherical Shell:\\ Newton's Method Using Implicit Coriolis Integration}

\author{Juan Cruz Gonzalez Sembla$^{1}$}
\author{Camille Rambert $^{1}$}, 
\author{Fred Feudel$^{2}$}
\author{Laurette S. Tuckerman$^{1}$}

\affiliation{
$^1${PMMH, CNRS, ESPCI Paris,} 
 {Universit\'e PSL, Sorbonne Universit\'e, Universit\'e de Paris,} 
 75005 Paris, France; {juan-cruz.gonzalez-sembla@espci.fr; camille.rambert@espci.fr; laurette@pmmh.espci.fr} 
\\
$^2$Institut f\"ur Physik und Astronomie, Universit\"at Potsdam, 
14476 Potsdam, Germany; {ffeudel@uni-potsdam.de}} 





\begin{abstract}
Geophysical flows are characterized by rapid rotation. Simulating these flows requires small timesteps to achieve stability and accuracy. Numerical stability can be greatly improved by the implicit integration of the terms that are most responsible for destabilizing the numerical scheme. We have implemented an implicit treatment of the Coriolis force in a rotating spherical shell driven by a radial thermal gradient. We modified the resulting timestepping code to carry out steady-state solving via Newton's method, which has no timestepping error. The implicit terms have the effect of preconditioning the linear systems, which can then be rapidly solved by a matrix-free Krylov method. We computed the branches of rotating waves with azimuthal wavenumbers ranging from 4 to 12. As the Ekman number (the non-dimensionalized inverse rotation rate) decreases, the flows are increasingly axially independent and localized near the inner cylinder, in keeping with well-known theoretical predictions and previous experimental and numerical results.
The advantage of the implicit over the explicit treatment also increases dramatically with decreasing $\Ek$, reducing the cost of computation by as much as a factor of 20 for Ekman numbers of order of $10^{-5}$. 
We carried out continuation for both the Rayleigh and Ekman numbers and obtained interesting branches in which the drift velocity remained unchanged between pairs of saddle--node bifurcations.
\end{abstract}

\maketitle






\section{Introduction}

{Convection in a rapidly rotating spherical shell subjected to differential heating or imposed heat flux and a radial gravity field is a standard model for planetary interiors. 
The onset of convection in this configuration has been extensively studied~\cite{chandrasekhar1961hydrodynamic,roberts1968thermal,busse1970thermal,zhang1987onset,jones2000onset,dormy2004onset,net2008onset,skene2024weakly}.
Due to the symmetry of the configuration, a~primary Hopf bifurcation leads to 
traveling (rotating) waves~\cite{ruelle1973bifurcations,ecke1992hopf,chossat_lauterbach00}, which take the form of fluid columns perpendicular to the axis of rotation, sometimes referred to as thermal Rossby waves~\cite{busse2002convective,hindman2022radial,dormy2025rapidly,potherat2025seven}. 
Scaling laws have been derived for the critical Rayleigh number and for the azimuthal wavenumber for these patterns and their drifting frequencies~\cite{dormy2004onset} as a function of the Ekman number, which is the ratio of the viscous to the Coriolis force. A~secondary Hopf bifurcation leads to modulated rotating waves, i.e.,\ quasi-periodic patterns with an additional modulation frequency~\cite{zhang1992convection,simitev_busse03,feudel2013multistability,feudel2015bifurcations,garcia2016continuation}.

{ 
Although convection in astrophysics follows the same basic fluid-dynamical principles as convection in geophysics, our work is primarily relevant to geophysical applications, in~which the density variations are sufficiently small that the fluid can be treated as Boussinesq, the~convection is due to temperature (rather than concentration) gradients, the~domain is a spherical shell, and~the Coriolis force is dominant.}
Concerning the physics of the problem, the~balance of forces can be used to define various regimes, each with a different convective length and velocity scales. In~particular, the Coriolis-inertial-buoyancy balance can be contrasted to a regime called viscous balance in which the viscous force replaces the inertial force~\cite{cardin1994chaotic,aubert2001systematic,jones2007thermal,aurnou2020connections}. 
Experimental and numerical studies have analyzed the effect of rotation on convection and on the generation of a magnetic field; these are reviewed in e.g.,~\cite{busse2002convective,potherat2025seven,jones2007thermal,olson2011laboratory}. Some notable numerical papers incorporating these effects are~\cite{glatzmaier1984numerical}, which describes a pioneering 3D code for solving the anelastic magnetohydrodynamic equations, and~\cite{christensen2001numerical} and~\cite{jones2011anelastic}, which compare the performance of various codes on Boussinesq and anelastic convection-driven dynamo benchmark problems, respectively (the anelastic approximation lies between the Boussinesq approximation and the fully compressible equations, capturing some compressibility effects while filtering out fast sound waves).
Laboratory experiments on radial gravity in a sphere are difficult to perform because they are affected by Earth's vertical gravity. For~this reason, microgravity experiments called Geoflow and Atmoflow, respectively, have been placed on a space station to mimic radial convection in rotating spherical annuli within the Earth~\cite{futterer2013sheet,Zaussinger_etal_2020} and in the atmosphere~\cite{TravnikovEgbers_2021}.

Three-dimensional numerical simulations in a spherical geometry usually represent the velocity and magnetic fields in terms of poloidal and toroidal potentials, which reduce a 3D solenoidal field to two scalar fields. These potentials and the temperature are decomposed into spherical harmonics in the angular directions, with~either Chebyshev polynomials or finite differences used in the radial direction.
Chebyshev collocation methods lead to dense matrices, but~sparse matrices can be obtained when using a Galerkin approach~\cite{tuckerman1989transformations,marti2016computationally}, resulting in lower time and memory~requirements.

Rapid rotation leads to a large Coriolis force.
The Coriolis force introduces coupling (of the velocity components and/or the spherical harmonics) and so it is usually treated explicitly in numerical codes. This leads to a significant limitation on the timestep. A~natural approach is then to treat this term implicitly, leading to a more stable scheme; its implementation for finite difference or Chebyshev collocation discretizations in the radial direction is discussed in~\cite{kuang1999numerical,hollerbach2000spectral}, respectively. The implicit treatment of the Coriolis term has been successfully implemented
 {by~\cite{sprague2006numerical,stellmach2008efficient,julien2024rescaled,van2024bridging} in a planar geometry and by~\cite{marti2016computationally, garcia2010comparison, garcia2016continuation, garciadormy2015}} in a spherical~geometry.

Another form of numerical computation is continuation, i.e.,\ the computation of steady states, traveling waves, or~other periodic states independently of their stability via Newton's or a related method.
Although unstable solutions cannot be observed experimentally, they are interesting because they form a framework for understanding the stable solutions, which in turn are accessible to experiment with. 

In this work, we describe the implementation and capabilities of a pair of numerical codes---two drivers relying on a common set of subroutines---that incorporate an implicit treatment of the Coriolis force to carry out both timestepping and also continuation. We will illustrate, test, and~compare the codes on the thermally driven rotating waves. 
We carried out numerical continuation in both Rayleigh and Ekman numbers. 
We found interesting examples of branches which contain saddle--node bifurcations separating plateaus in drift frequency.
Our numerical algorithm uses the basic framework described in~\cite{hollerbach2000spectral,feudel2011convection,feudel2013multistability,feudel2015bifurcations,feudel2017hysteresis} and the spherical harmonic library of~\cite{schaeffer2013efficient}.
The research which is closest to ours is that of~\cite{sanchez2004newton,garcia2008antisymmetric,sanchez2013computation,garcia2016continuation}, who, like us, have carried out implicit timestepping and continuation for convection in a rotating spherical shell and explored various interesting parameter regimes. There are a number of differences between our continuation method and theirs, some of which have been analyzed { in}~\cite{tuckerman2018order} and which we will discuss after we have described our numerical~methods.

In Section~\ref{sec:physical_description}, we describe the problem, governing equations and non-dimensional parameters in addition to introducing
rotating wave solutions
of different azimuthal symmetries. The~numerical methods, including spatial discretization, Newton, and path-following methods are detailed in Section~\ref{sec:numerical_methods}. Continuation results are described in Section~\ref{sec:branch_following}, while timing comparisons and timestepping results are  presented in Section~\ref{sec:timing_comparisons}.
We discuss and provide a conclusion to our results in Sections~\ref{sec:discussion} and \ref{sec:conclusion}.
}    



\section{Physical~Description}
\label{sec:physical_description}

\subsection{Governing~Equations}

\begin{figure}
  \centering
  \includegraphics[width=0.5\columnwidth]{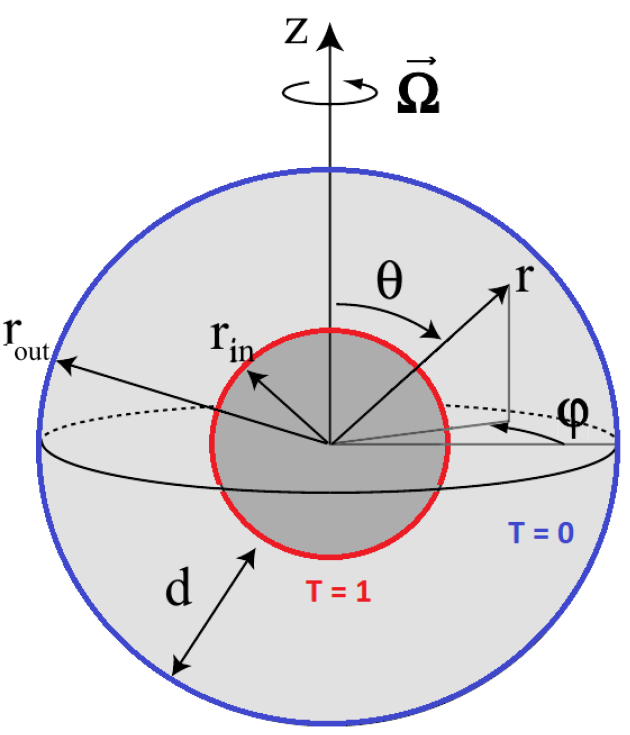}
\caption{{Geometry.} 
 The~domain is the spherical annulus between an inner and outer sphere of radius $r_{\rm in}$ and $r_{\rm out}$ whose temperatures are fixed at $T=1$ and $T=0$. Both spheres rotate around the $z$ axis with angular velocity $\Omega$. 
Figure taken and adapted from \cite{magic_figure}.}
\label{fig:geometry}
\end{figure}

We study classical Rayleigh--B\'enard convection in a spherical fluid shell rotating with constant angular velocity $\Omega$ about the $z$ axis, as illustrated in Figure \ref{fig:geometry}.
The~shell is heated from within by imposing a temperature difference $\Delta T$ between the inner and outer spherical boundaries. Lengths are non-dimensionalized with the gap size $d$, so that the dimensionless outer and inner radii are $r_{\rm out}$ and $r_{\rm in} = r_{\rm out} - 1$, respectively. The~aspect ratio can be specified via $r_{\rm out}$ or via $\eta \equiv r_{\rm in}/r_{\rm out} = (r_{\rm out}-1)/r_{\rm out}$. Time is scaled by the viscous diffusion time $d^2/\nu$, where $\nu$ is the kinematic viscosity and the velocity is scaled by the viscous diffusion velocity $\nu/d$.
Measuring temperature from a reference temperature $T_0$ at which the mass density is $\rho_0$, we non-dimensionalize pressure by $\rho_0\nu$, and~temperature by $\Delta T$.

The gravitational acceleration $g$ is assumed to be proportional to the distance $r$ from the center of the sphere (as is the case for self-gravity in the interior of a spherical body with constant mass density) and is thus of the form $g = -(g_o /r_{\rm out})r$, where $g_o$ is the acceleration at radius $r_{\rm out}$. The~resulting non-dimensional Boussinesq equations {read}

\begin{subequations}
\begin{align}
\Ek\left[\frac{\pd \bU}{\pd t} + \left(\bU \cdot \nabla\right) \bU -\nabla^2\bU\right] &=-\nabla P + 
\Ra\; \Temp \:\frac{\br}{r_{\rm out}} 
- 2\ez \times \bU
   \label{eq:Navier-Stokes}\\
\frac{\pd \Temp}{\pd t} + \bU \cdot \nabla \Temp &= \frac{1}{\Pra} \lap T 
   \label{eq:temp}\\
   \divv \bU &= 0  
   \label{eq:div}
\end{align}
\end{subequations}
When the density is constant, the~gravitational and centrifugal forces
can be written as gradients and included in the pressure
gradient. Variable density leads to a non-gradient portion of these
forces. The~non-gradient part arising from gravity drives the
convection. We make the common assumption in the geophysics of
neglecting the non-gradient part arising from the centrifugal
force~\cite{sanchez2013computation,potherat2025seven};
see~\cite{marques2007centrifugal,satake2022influence,potherat2025seven}
for examples of the effect of centrifugal~buoyancy.

The non-dimensional parameters used in Equation~\eqref{eq:Navier-Stokes} above are the Ekman, Prandtl, and~Rayleigh numbers:
\begin{align}
    \Ek\equiv\frac{\nu}{\Omega d^2}, \quad \Pra\equiv\frac{\nu}{\kappa}, \quad \Ra = {\Ra}_{\rm rot}\equiv\frac{d \alpha g_o \Delta T}{\Omega\nu}
    \end{align}
    where $\kappa$ is the thermal diffusivity and $\alpha$ is the thermal expansion coefficient.
    The Rayleigh number used here is adapted to the rotating case and is related to the conventional thermal Rayleigh number $\Ra_{\rm therm}$ by
\begin{equation}    
    \Ra_{\rm therm} \equiv \frac{d^3\alpha g_0 \Delta T}
    {\kappa\nu}= \frac{d \alpha g_o \Delta T}{\Omega\nu} \:\frac{\Omega d^2}{\nu}\:\frac{\nu}{\kappa} = \Ra_{\rm rot} \:\frac{\Pra}{\Ek}
    \end{equation}   
    {In what}  follows, we will denote $\Ra_{\rm rot}$ merely by $\Ra$. 
      No-slip and fixed-temperature boundary conditions are applied at the inner and outer radii:
      \begin{subequations}
      \label{eq:BCs}
\begin{align}
     \bU(r_{\rm in})=0 &\quad \bU(r_{\rm out})=0 \label{eq:BCvel}\\
      T(r_{\rm in})=1 &\quad T(r_{\rm out})=0\label{eq:BCtemp}
\end{align}
      \end{subequations}
and the conductive solution is
\begin{align}
\bU_{\rm cond}=0, &\quad T_{\rm cond}(r) = \frac{r_{\rm out} \; r_{\rm in}}{r} - r_{\rm in}
\end{align}
{Following} the benchmark study~\cite{christensen2001numerical}, we set the Prandtl number to $\Pra=1$ and the radii to $r_{\rm out}=20/13$ and $r_{\rm in}=7/13$ (so that $\eta=7/20=0.35$) throughout the investigation and we will vary $\Ek$ and $\Ra$.

\subsection{Overview}
\label{sec:overview}

\begin{figure}
\subfloat[\centering]
{\includegraphics[height=6cm]{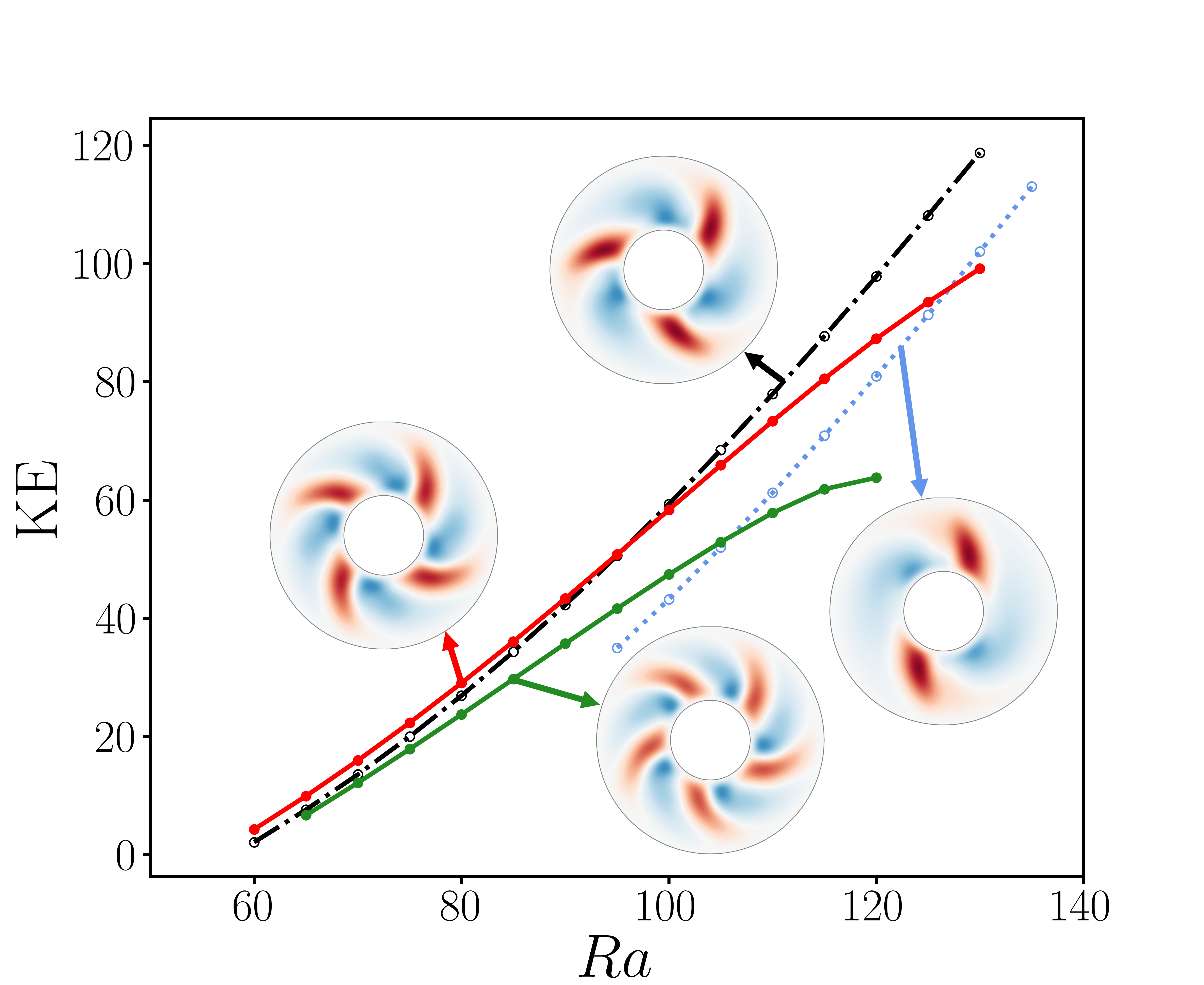}}
\subfloat[\centering]
{\includegraphics[height=6cm]{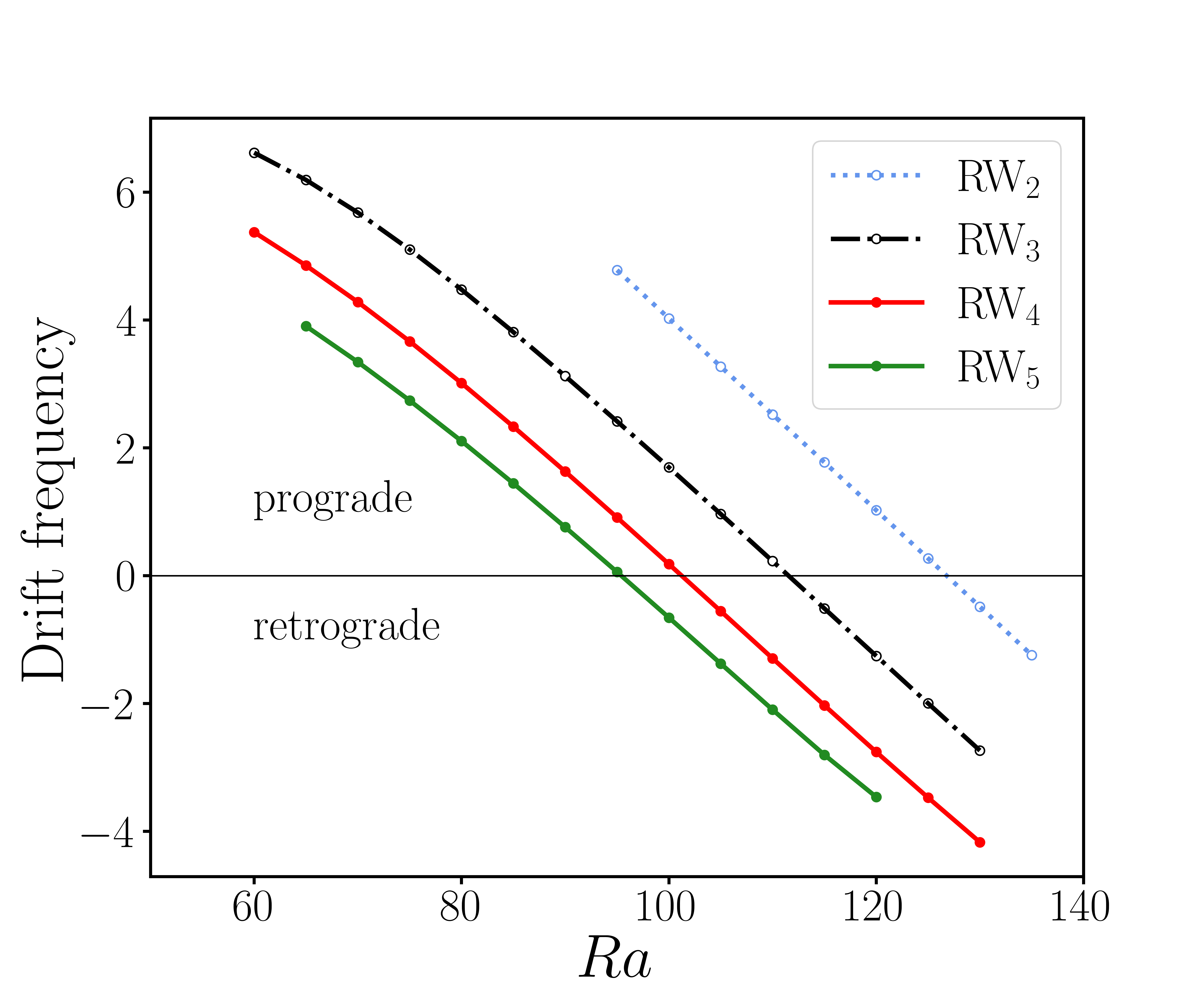}}
\caption{(\textbf{a}) Bifurcation 
 diagram of convection in a rotating spherical annulus for $\Ek=10^{-3}$. Branches RW$_2$ (blue), RW$_3$ (black), RW$_4$ (red), and RW$_5$ (green) are shown.  (\textbf{a}) Kinetic energy density. (\textbf{b}) Drift frequency. The~drift frequencies for each rotating wave decreases from prograde (faster than imposed velocity $\Omega$) to retrograde (slower than $\Omega$) with increasing $\Ra$ with the same slope. The~resolution used for these simulations was $(N_r, N_\theta, N_\phi \times M) = (46, 72, 128 \times 1)$.}
\label{fig:Feudel}
\end{figure}

The first states to appear at the onset of convection are rotating waves. We denote these by RW$_M$, where $M$ is the azimuthal wavenumber of the rotating wave. 
Figure~\ref{fig:Feudel}a,b, modeled on those in~\cite{feudel2013multistability}, display properties of some of these rotating wave solutions for $\Ek=10^{-3}$. Four traveling waves are shown, with~azimuthal wavenumbers from $M=2$ to $M=5$. 
At $\Ek=10^{-3}$, RW$_4$ is the first in Rayleigh number to bifurcate, and~therefore, it is the only one which is stable at onset. The~drift frequency, i.e.,\ the frequency relative to the imposed frequency $\Omega$, decreases as $\Ra$ increases, from~prograde (faster than $\Omega$) to retrograde (slower than $\Omega$). These results were obtained via timestepping simulations starting from the initial conditions of the form $\cos(M \phi)$.

\begin{figure}
\includegraphics[width=\textwidth]
{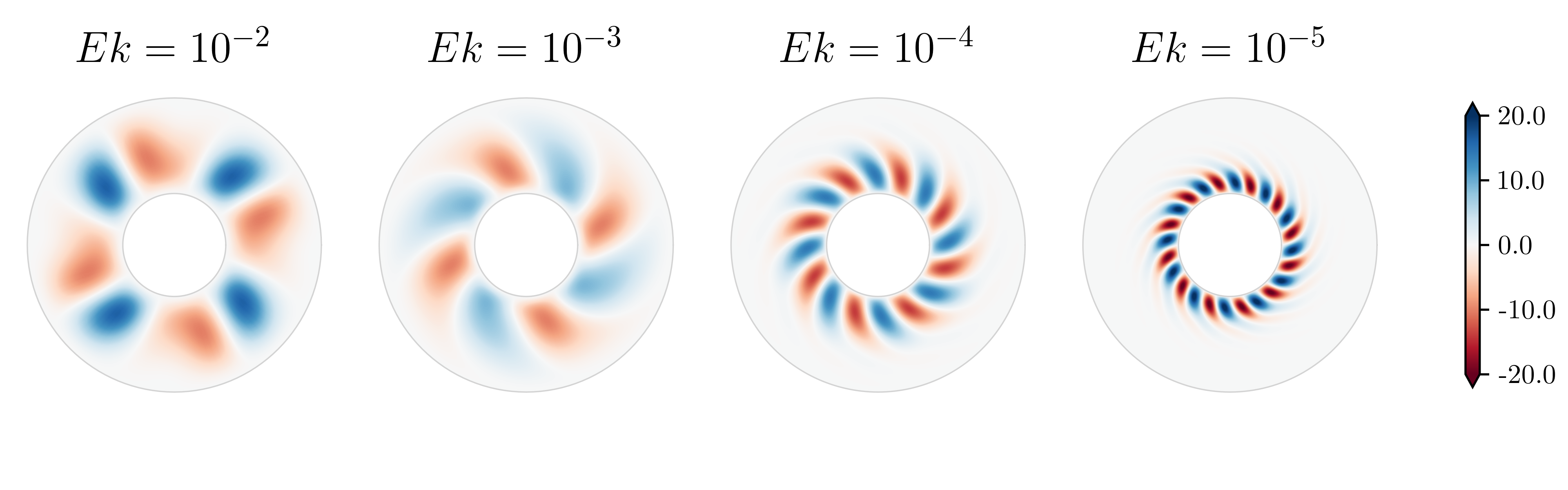}
\includegraphics[width=\textwidth]
{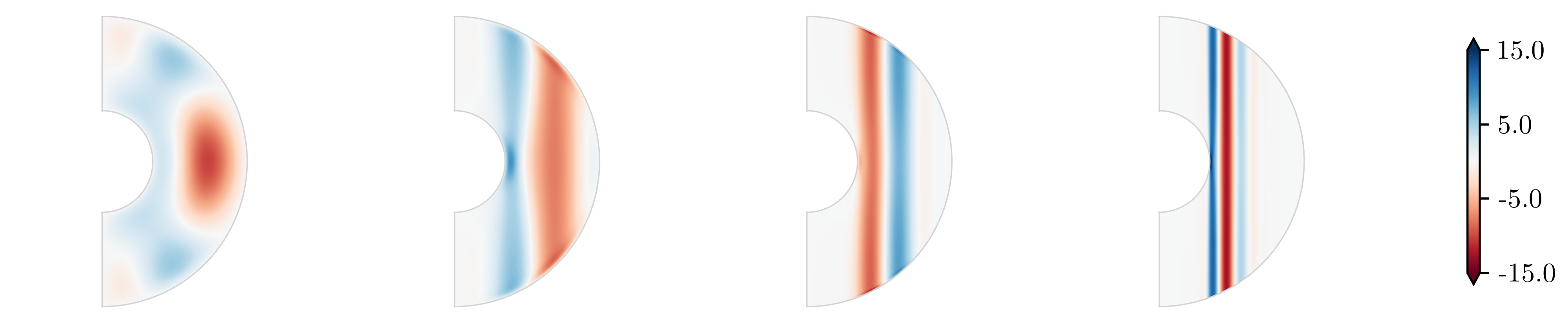}
\caption{{Velocity fields} 
 of rotating waves as the Ekman number is reduced. The above shows RW$_4$ for $\Ek~=~10^{-2}$ and $Ra = 140$; RW$_4$ for $\Ek = 10^{-3}$ and $Ra = 77$; RW$_8$ for $\Ek = 10^{-4}$ and $Ra = 88$; and RW$_{12}$ for $\Ek = 10^{-5}$ and $Ra = 126$. { A state with azimuthal wavenumber $M$ has $M$ convection cells, forming $M$ Taylor columns.} All of these states have an approximate kinetic energy density of 25. For~each field, the~radial velocity $U_r$ on the equatorial plane (\textbf{top}) and the azimuthal velocity $U_\phi$ on a meridional plane (\textbf{bottom}) are shown. As~$\Ek$ decreases, the~convective region becomes more confined near the inner sphere and the flow becomes almost axially independent, according to the Taylor--Proudman theorem. The~resolutions used for these simulations were \mbox{$(N_r, N_\theta, N_\phi \times M) = (46, 72, 32 \times 4)$} for RW$_4$ at both $\Ek = 10^{-2}$ and $\Ek = 10^{-3}$; $(60, 100, 28 \times 8)$ for RW$_8$; and $(90, 184, 32 \times 12)$ for RW$_{12}$.}
\label{fig:four_rotating_waves}
\end{figure}

\begin{figure}
\includegraphics[width=0.9\textwidth]{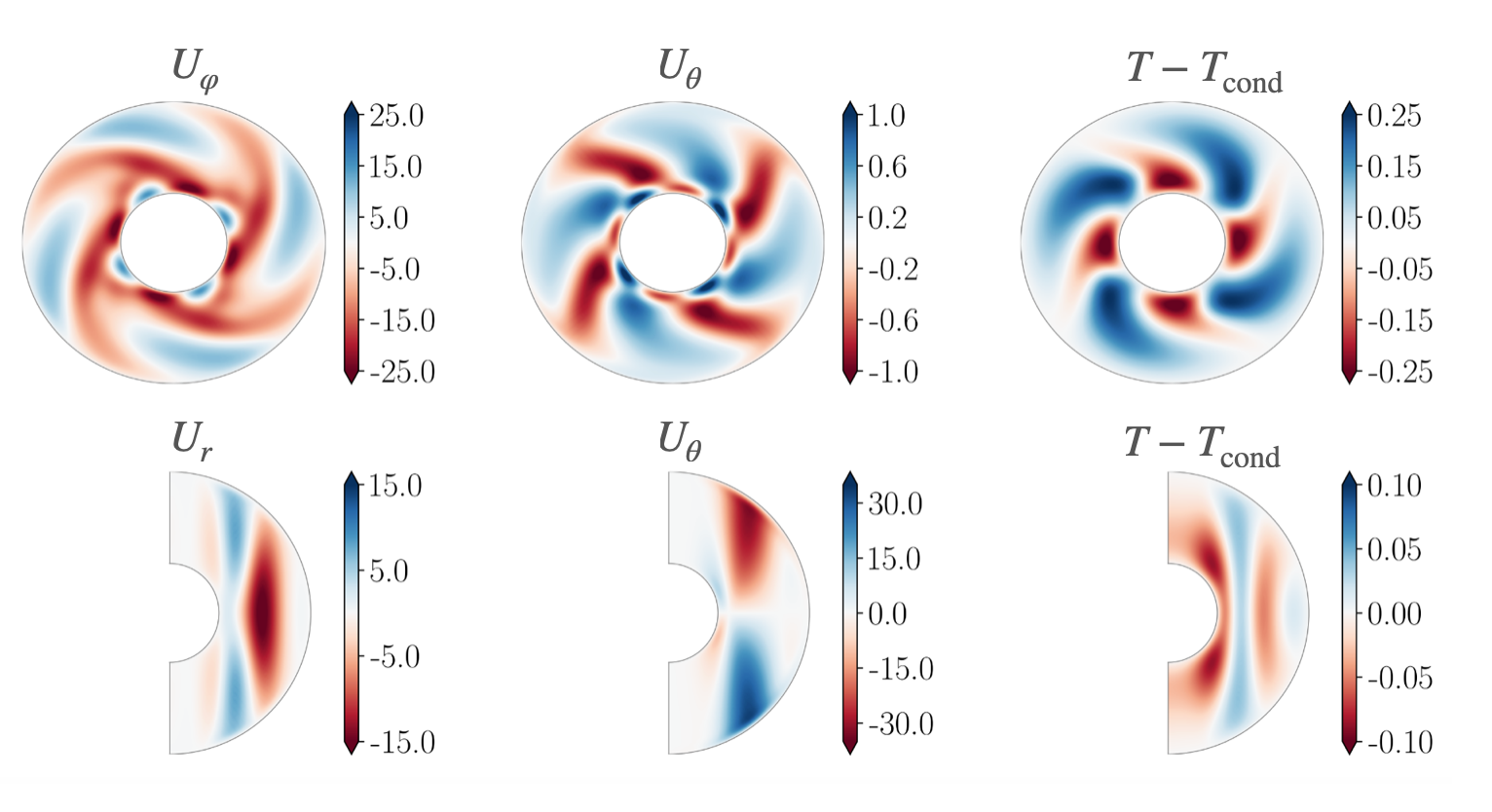}
\caption{
Visualizations
 of the components of RW$_4$ at $Ek = 10^{-3}$ and $Ra = 140$. \textbf{Top}: equatorial plane. \textbf{Bottom}: meridional plane.
The resolution used here was $(N_r, N_\theta, N_\phi \times M) =~(46, 72, 32 \times 4)$.}
\label{fig:flow_RW4_Ek_1e-3}
\end{figure}

Figure~\ref{fig:four_rotating_waves} shows the qualitative evolution of the rotating waves as $\Ek$ is decreased from $10^{-2}$ to $10^{-5}$: the flow is increasingly stronger near the inner sphere and the azimuthal wavenumber increases. These two properties are related: the radial interval over which the convection is most active decreases, and the azimuthal wavelength decreases as well \cite{dormy2004onset,garciadormy2015}. RW$_8$ appeared naturally when performing timestepping simulations from an initial condition of the form $\cos(4\phi)$ at $\Ek = 10^{-4}$, as~did RW$_{12}$ at $\Ek = 3 \times 10^{-5}$. This result was then used as input for a path-following computation to obtain RW$_{12}$ at $\Ek = 10^{-5}$. The~same techniques were applied to obtain RW$_4$ at $\Ek = 10^{-2}$ starting from RW$_4$ at $\Ek = 10^{-3}$.
Figure \ref{fig:flow_RW4_Ek_1e-3} gives a more detailed visualization of an RW$_4$ state.

\unskip

\section{Numerical~Methods}
\label{sec:numerical_methods}
\unskip
\subsection{Spatial~Representation}
The velocity is decomposed into toroidal and poloidal potentials $e$ and $f$ such that
\begin{equation}
    \bU(r,\theta,\phi,t) = \curl e \:\be_r + \curl\curl f \be_r
\end{equation}
These potentials and the temperature are expanded in spherical
harmonics $Y^{m}_{l}(\theta,\phi)$ and Chebyshev polynomials $T_k$ of
the shifted radius
$x\equiv(2r-(r_{\rm out}+r_{\rm in}))/(r_{\rm out}-r_{\rm in})$.%
\begin{equation}
    f(r, \theta, \phi, t) = \sum_{k, l, m} f_{k, l, m}(t) T_{k} \left( x \right) Y_l^m(\theta, \phi)
    \label{eq:expansion}
\end{equation}

To carry out transformations to and from the spherical harmonic space,
we use the {\tt {SHTns} } library~\cite{schaeffer2013efficient}. This
is a highly efficient code that uses the {\tt FFTW} library to perform
fast transformations to and from Fourier space, and~the recursion
relations presented in~\cite{ishioka2018} for an optimized computation
of the associated Legendre polynomials. Although~the library allows
for multithreaded computations as well as GPU offloading, these
capabilities were not used in the present work.  { This library has
  promising scalability prospects for computations in massively
  parallel clusters, as~demonstrated in the {\tt XSHELLS} and {\tt
    MagIC} codes~\cite{schaeffer2017turbulent,
    lago2021magic}. Furthermore, our code agrees with the non-magnetic
  benchmark proposed in~\cite{christensen2001numerical} and the
  results presented in~\cite{feudel2013multistability}, validating our
  numerical method as well as our choice of using the {\tt SHTns}
  library.  }

De-aliasing in the $r$ direction is carried out according to the $3/2$ rule, whereas in the $\theta$ and $\phi$ directions, it is performed by {\tt SHTns}. To~achieve this, the~library determines a suitable number of physical points based on the nonlinear order of the equations to be solved (two in this case) with the aim of maximizing the efficiency of the FFT algorithm. Moreover, for~solutions which have azimuthal periodicity $2\pi/M$, we restrict our calculations to a segment of the sphere $0\leq \phi < 2\pi/M$, or, equivalently, to~those whose Fourier expansions contain only multiples of the wavenumber $M$; this is easily implemented using options in {\tt SHTns}. Lastly, we will denote the number of gridpoints in $r, \theta$ and $\phi$ as $N_r,  N_\theta$ and $N_\phi \times M$, respectively. 
The resolutions chosen for the simulations are similar to those that have been shown to be adequate in the benchmark paper~\cite{christensen2001numerical} and in more recent studies~\cite{net2008onset, garciasancheznet2014, garcia2016continuation}. The~spatial resolutions are given in the captions of each of the figures.

The radial component of the curl and of the double curl of Equation \eqref{eq:Navier-Stokes} are taken, leading to
\begin{subequations}
\begin{align}
     &\sum_{l, m} \frac{l(l+1)}{r^2} \Ek \left( \frac{\pd}{\pd t} - L_{l} \right) e_{l, m}(r,t)  Y_l^m (\theta,\phi)= \boldsymbol{e_{r}} \cdot \curl \bF 
\label{eq:expcor_e}\\
- &\sum_{l, m} \frac{l(l+1)}{r^2} \Ek \left(\frac{\pd}{\pd t} -  L_{l} \right) L_{l} f_{l, m}(r,t) Y_l^m (\theta,\phi) = \er \cdot \curl \curl \bF 
     \label{eq:expcor_f}
     \end{align}
     \end{subequations}
where
\begin{align}
L_l \equiv\frac{\pd^2}{\pd r^2} - \frac{l(l+1)}{r^2} \qquad\mbox{and} \qquad
\bF  \equiv \Ek \;(\bU\cdot\nabla)\bU - \nabla P + \frac{Ra}{r_{\rm out}}\; T \;\boldsymbol{r}- 2\ez \times \bU
\label{eq:defineFexp}
\end{align}
Equations \eqref{eq:expcor_e} and \eqref{eq:expcor_f} are decoupled in
$\ell$, $m$ and in potential ($e$ vs. $f$), and~so their discrete
versions involve block-diagonal matrices, with an $N_r \times N_r$
block for each $(l,m)$ pair, as in Figure \ref{fig:blocks}(a).
Boundary conditions \eqref{eq:BCvel} on $\bU$ become
\begin{align}
e_{lm}= f_{lm} = df_{lm}/dr =0 \mbox{~~ at~} r_{\rm in}, r_{\rm out}
\end{align}
The number of boundary conditions on $e$ and $f$ matches the fact that
Equation \eqref{eq:expcor_e} for $e$ is of second order in $r$ while
Equation \eqref{eq:expcor_f} for $f$ is of fourth order in
$r$. The~boundary conditions are imposed by replacing the last two
rows of Equation \eqref{eq:expcor_e} or the last four rows of
Equation~\eqref{eq:expcor_f}, corresponding to the highest Chebyshev
polynomials. See,
for~example,~\mbox{\cite{gottlieb1977numerical,hollerbach2000spectral}}.

The temperature field is also expanded in Chebyshev polynomials and spherical harmonics like \mbox{Equation~\eqref{eq:expansion}}. Its evolution is governed by the discretized form of \mbox{Equation \eqref{eq:temp}} and the boundary conditions in Equation \eqref{eq:BCtemp} on the temperature are imposed straightforwardly on the rows corresponding to the last two Chebyshev~polynomials.

\subsection{Implicit Coriolis~Integration}
\label{sec:ImpCor}

To carry out the implicit integration of the Coriolis term~\cite{hollerbach2000spectral}, we include the Coriolis force in the left-hand-side and remove it from ${\bf F}$, leading to
\begin{subequations}
\begin{align}
   & \sum_{m,l} \left [ \frac{l(l+1)}{r^2} \Ek \left( \frac{\pd}{\pd t} - L_{l} \right) -\frac{2im}{r^2} \right] e_{lm}(r,t) Y_l^m(\theta,\phi) \nonumber\\
     &+ \frac{2}{r^2} \left( \frac{l(l+1)}{r}-\frac{\pd}{\pd r} \right) f_{lm}(r,t) \sin{\theta} \frac{d}{d\theta} Y_l^m(\theta,\phi) \nonumber\\
    & + 2\frac{l(l+1)}{r^2} \left(\frac{2}{r}-\frac{\pd}{\pd r} \right) f_{lm}(r,t) \cos{\theta}  \: Y_l^m(\theta,\phi) = \boldsymbol{e_{r}} \cdot \curl \bF^{\rm implicit}
     \label{eq:impcor_e_init}
     \end{align}
\begin{align}
  &  \sum_{m,l} - \left [ \frac{l(l+1)}{r^2} \Ek \left(\frac{\pd}{\pd t} - L_{l} \right) -\frac{2im}{r^2} \right] L_{l} f_{\rm lm}(r,t) Y_{l}^m(\theta,\phi) \nonumber \\
   &  + \frac{2}{r^2} \left( \frac{l(l+1)}{r}-\frac{\pd}{\pd r} \right) e_{lm}(r,t) \sin{\theta} \frac{d}{d\theta} Y_{l}^{|m|}(\theta,\phi)\nonumber \\
   &  + 2\frac{l(l+1)}{r^2} \left(\frac{2}{r}-\frac{\pd}{\pd r} \right) e_{lm}(r,t) \cos{\theta}  \:Y_{l}^m(\theta,\phi) = \boldsymbol{e_{r}} \cdot \curl \curl \boldsymbol{F}^{\rm implicit}
      \label{eq:impcor_f_init}
\end{align}
    \end{subequations}
      where
\begin{align}
 \qquad
\bF^{\rm implicit}  \equiv \Ek \;(\bU\cdot\nabla)\bU - \nabla P + \frac{\Ra}{r_{\rm out}}\; T \;\boldsymbol{r}
\label{eq:defineFimp}
\end{align}

The recursion relations of the Legendre polynomials
\begin{subequations}
\label{eq:recu_all}
\begin{align}
    \sin{\theta} \frac{d}{d\theta} P_{l}^{|m|} &= \frac{l(l-|m|+1)}{2l+1}P_{l+1}^{|m|} - \frac{(l+|m|)(l+1)}{2l+1}P_{l-1}^{|m|} 
    \label{eq:recu1}\\
    \cos{\theta} P_{l}^{|m|} &= \frac{l-|m|+1}{2l+1}P_{l+1}^{|m|} + \frac{(l+|m|)}{2l+1}P_{l-1}^{|m|} 
    \label{eq:recu2}
\end{align}
\end{subequations}
together with
\begin{align}
Y_l^m(\theta,\phi) = N_l^m P_l^m(\cos\theta) e^{im\phi} \qquad\qquad
N_l^m=\sqrt{(2-\delta_{m0})(2l+1)\frac{(l-m)!}{l+m)!}}
\label{eq:Ylm_norm}\end{align}
can be used to transform Equations \eqref{eq:impcor_e_init} and \eqref{eq:impcor_f_init} into 
%
\begin{subequations}
\begin{align}
  &  \sum_{m,l} 
    Y_l^m(\theta,\phi)
    \left\{\left [ \frac{l(l+1)}{r^2} \Ek \left( \frac{\pd}{\pd t} - L_{l} \right) -\frac{2im}{r^2} \right] e_{lm}(r,t)\right.\nonumber\\
  &   + \frac{2}{r^2} \frac{(l-1)(l+1)(l-|m|)}{2l-1}  \left [  \frac{l}{r}-\frac{\pd}{\pd r}   \right]\frac{N_{l-1}^m}{N_l^m} f_{l-1,m}(r,t)
    \nonumber  \\
  &  \left. + \frac{2}{r^2} \frac{l(l+2)(l+1+|m|)}{2l+3} \left [ -\frac{(l+1)}{r}-\frac{\pd}{\pd r} \right] \frac{N_{l+1}^m}{N_l^m}f_{l+1,m}(r,t)\right\}
=\boldsymbol{e_{r}} \cdot \curl \bF^{\rm implicit} 
\label{eq:impcor_e_final}
\end{align}
\begin{align}
&\sum_{m,l} Y_l^m(\theta,\phi)\left\{ - \left [ \frac{l(l+1)}{r^2} \Ek \left( \frac{\pd}{\pd t} - L_l \right) -\frac{2im}{r^2} \right] L_lf_{lm}(r,t)\right. \nonumber\\
&     + \frac{2}{r^2} \frac{(l-1)(l+1)(l-|m|)}{2l-1}  \left [ \frac{l}{r}-\frac{\pd}{\pd r}  \right] \frac{N_{l-1}^m}{N_l^m} e_{l-1,m}(r,t)\nonumber
     \\
 &   \left.+ \frac{2}{r^2} \frac{l(l+2)(l+1+|m|)}{2l+3} \left [ -\frac{(l+1)}{r}-\frac{\pd}{\pd r}  \right]  \frac{N_{l+1}^m}{N_l^m} e_{l+1,m}(r,t)\right\}
 =\boldsymbol{e_{r}} \cdot \curl \curl \bF^{\rm implicit} 
 \label{eq:impcor_f_final}
\end{align}
\end{subequations}
Appendix A  presents further details of these computations.
In contrast with\linebreak \mbox{Equations \eqref{eq:expcor_e} and \eqref{eq:expcor_f}}, Equations \eqref{eq:impcor_e_final} and \eqref{eq:impcor_f_final} are coupled in several~ways:
\begin{itemize}
\item The spectral coefficients $e_{\rm lm}$ and $f_{\rm lm}$ both appear in both~equations.

\item While each $m$ can be treated independently, components $\ell$, $\ell+1$, and~$\ell-1$ are~coupled.

\item The real and imaginary parts of $e_{\rm lm}$ and $f_{\rm lm}$ are coupled via
the imaginary coefficient $2im/r^2$.

\end{itemize}

\begin{figure}
\subfloat[\centering]{\includegraphics[width=6cm]{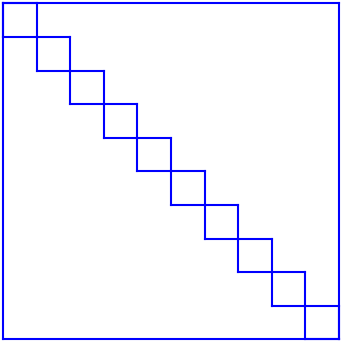}}
\hfill
\subfloat[\centering]{\includegraphics[width=6cm]{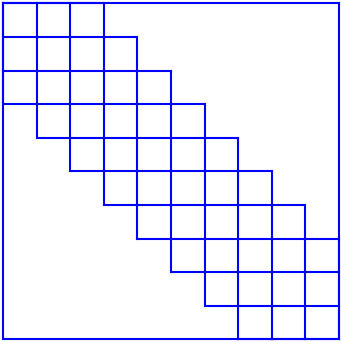}}
\caption{(\textbf{a}) Block-diagonal matrix. (\textbf{b}) Block-pentadiagonal~matrix.\label{fig:blocks}}
\end{figure}

However, two decoupled classes of coefficients appear, with one class containing coefficients with odd $l$ for $e_l$ and even $l$ for $f_l$ and a second class with the opposite property. 
For example, in Equation \eqref{eq:impcor_e_final}, the~real component of $e_{lm}$ is coupled to its imaginary component and to the real parts of $f_{l\pm 1,m}$ (with opposite parity to $l$) and in Equation \eqref{eq:impcor_f_final}, the~real component of $f_{lm}$ is coupled to its imaginary component and to the real parts of $e_{l\pm 1,m}$.
For each class, the~sums in Equations \eqref{eq:impcor_e_final} and \eqref{eq:impcor_f_final} are represented by a block pentadiagonal matrix, as~illustrated in Figure~\ref{fig:blocks}b. 

\subsection{Newton~Method}
\label{sec:Newton}

We represent our system of governing equations schematically by
\begin{align}
\frac{\pd \bU}{\pd t}= \mN(\bU) + \mL \bU
\label{eq:ode}
\end{align} 
For our problem, $\bU$ corresponds to the poloidal and toroidal fields $e$ and $f$ and the temperature $\Temp$. For~explicit or implicit Coriolis integration, $\mN(\bU)$ corresponds to the right-hand-sides of Equations \eqref{eq:expcor_e} and \eqref{eq:expcor_f} or of Equations  \eqref{eq:impcor_e_init} and \eqref{eq:impcor_f_init}, respectively. 
$\mL\bU$ corresponds to minus the left-hand-sides of the same equations with the time derivative omitted (the correspondence is imperfect, since $\pd/\pd t$ does not act directly on $\bU$.)

We represent our timestepping code schematically by the following implicit--explicit Euler method:
\begin{align}
 \bU(t+\dt) &= \bU(t) + \dt(\mN(\bU(t))+\mL \bU(t+\dt)\nn)\\
\Longrightarrow \bU(t+\dt) &= \mB(\bU(t)) \equiv (I - \dt \mL)^{-1}\left(I + \dt\mN\right)\bU(t)
 \label{eq:discrete}
\end{align}
where $\mL$ is linear (though $\mN$ does  not need to be nonlinear).
Steady states are solutions to
\begin{align}
0=\mA(\bU) \equiv\mN(\bU)+\mL\bU
\label{eq:steady}
\end{align}
or, alternatively
\begin{align}
    0=\bU(t+\dt)-\bU(t)
    \label{eq:continuous}
\end{align}
where $\bU$ is a solution to the continuous-time differential Equation
\eqref{eq:ode}. Surprisingly, Equation \eqref{eq:continuous} is also a
criterion for stationarity when $\bU(t+\dt)$ is calculated via the
timestepping scheme Equation \eqref{eq:discrete} {\it {for any value
    of} $\dt$}, as~can be seen from
\begin{align}
(\mB-I)(\bU(t))=\bU(t+\dt) -\bU(t) &= (I - \dt \mL)^{-1}\left(I + \dt\mN\right)\bU(t)-\bU(t)\nn\\
 &= (I - \dt \mL)^{-1}\left(I + \dt\mN-(I-\dt\mL))\bU(t)\right)\nn\\
 &= \underbrace{(I - \dt \mL)^{-1} \dt}_{\textstyle\mP(\dt)}\underbrace{(\mN+\mL)}_{\textstyle\mA}\bU(t)\label{eq:finalrhs}
 \end{align}
Thus, $\mB-I=\mP(\dt)\mA$ and so
roots of $\mA$ of Equation \eqref{eq:steady} can be found by computing roots of $\mB-I$, which is simply the difference between two large consecutive timesteps.
Note that Equation \eqref{eq:finalrhs} does not use Taylor series; $\dt$ can be of any size. Indeed, for~$\dt$ large, $\mP(\dt) \rightarrow -\mL^{-1}$ acts as a preconditioner (approximate inverse) for $\mA$, whose poor conditioning is due to that of $\mL$. 
 
Finding the roots of Equation \eqref{eq:finalrhs} via Newton's method requires the solution of the linear~system
\begin{align}
(\mB_{\bU}-I)\bu&=(\mB-I)\bU
\label{eq:precond}
\end{align}
where $\bU$ is the current estimate and $\mB_\bU$ is the Jacobian of $\mB$ evaluated at $\bU$.
{ 
The linear Jacobian operator $\mB_\bU$ acting on $\bu$ is derived from $\mB$ by modifying the nonlinear operator $\mN$.
We recall that $\bU$ represents $(\bU,T)$ and $\bu$ represents $(\bu,\tau)$ where $\tau$ is a temperature perturbation.
To form $\mN_\bU$, we replace the nonlinear terms in Equations \eqref{eq:Navier-Stokes} and \eqref{eq:temp} as~follows:
\begin{subequations}
\begin{align}
(\bU\cdot\grad)\bU&\Longrightarrow (\bU\cdot\grad)\bu + (\bu\cdot\grad)\bU\\
\bU\cdot\grad T &\Longrightarrow \bU\cdot\grad\tau + \bu\cdot\grad T
\end{align}
\end{subequations}
These substitutions produce $\mN_\bU$, which is then used in Equation
\eqref{eq:discrete} to form $\mB_\bU$.  In addition, homogeneous
boundary conditions must be applied to the perturbations $\bu$ and
$\tau$ instead of the inhomogeneous boundary conditions Equation
\eqref{eq:BCs}.  We emphasize that the operations on both sides of
Equation \eqref{eq:precond} are just the ordinary and the linearized
timestepping operators $\mB$ and $\mB_\bU$.}

We solve Equation \eqref{eq:precond} using a Krylov method such as GMRES~\cite{Saad_Schultz86}. In~such matrix-free methods, one only needs to provide the right-hand-side $(\mB-I)\bU$ for the current estimate, and~a routine which carries out the {\it {action}}
of $(\mB_\bU-I)$ on any vector $\bu$. The~computational cost of a Newton step can be measured by the number of such actions (i.e.,\ GMRES iterations) taken by the Krylov method, since the right-hand-side remains constant throughout the step. The~number of GMRES iterations required is low if  $(\mB_\bU-I)=\mP(\dt)\mA$ is well conditioned; it is for this reason that we take $\dt$ large. Once the decrement $\bu$ is determined by solving  Equation~\eqref{eq:precond}, it is subtracted from $\bU$ to form an improved estimate. 
We accept $\bu$ as the decrement if the linear system Equation \eqref{eq:precond} is solved by GMRES to relative accuracy $10^{-10}$. For~Newton's method, we accept $\bU$ as a steady state, i.e.,  a solution to Equation~\eqref{eq:continuous}, if~$||(\mB-I)\bU||<10^{-7}$.
{
Similar values of this tolerance have been used in~\cite{garcia2016continuation}. In~addition, solutions found via Newton's method are far more accurate than those found by timestepping, which is usually halted when a solution remains constant to three to five significant digits.
}



{The method described above is called \emph{{Stokes preconditioning}}.
  It was first applied to calculate bifurcation diagrams in spherical Couette flow~\cite{tuckerman1989steady,mamun1995asymmetry} and subsequently to a wide variety of hydrodynamic problems, e.g.,\  
\cite{bergeon1998marangoni,nore20031,meca2004blue,batiste2006spatially,boronska2010extreme_II,mercader2011convectons,beaume2013convectons,torres2014bifurcation} 
and even to Bose--Einstein condensation~\mbox{\cite{huepe1999decay,huepe2003stability}}. In~Stokes preconditioning, $\dt$ no longer plays the role of a timestep but serves to better condition the system of linear equations in each Newton step. We investigated the effect of the choice of $\dt$ on Newton solving with Stokes preconditioning in~\cite{tuckerman1989steady,tuckerman2018order} and on the computation of eigenvalues by a similar method in~\cite{tuckerman2015laplacian}. 
In these articles, time has been non-dimensionalized by the advective time, and~the usual value used in timestepping is $\dt\approx 0.01$.
An increase in the value of $\dt$ from $0.01$ to $0.1$ and then to $10$ leads to a decrease in the required CPU-time for convergence GMRES by \mbox{1--2 orders} of magnitude per $\dt$ decade until an asymptotic limit of $\dt \approx 100$ or 1000 is reached. We emphasize that the final result is independent of the choice of $\dt$: only the time taken for convergence is affected. In~the current study, we use the value $\dt=200$ for Newton~solving.}

The name of the method comes from the fact that the linear operator $\mL$ which is integrated implicitly in Equation \eqref{eq:discrete} is usually the viscous term that occurs in Stokes equation. 
{ This preconditioning becomes less powerful as the Reynolds number increases and other terms begin to dominate the viscous term; this effect was demonstrated and measured quantitatively  in~\cite{tuckerman2018order}.} 
Hence, it is beneficial to include as many possible other terms in $\mL$, as~long as they are linear and can be inverted directly. 
It is for this reason that we have included the Coriolis force in $\mL$, using the equations derived in Section~\ref{sec:ImpCor}.
In Section~\ref{sec:timing_comparisons}, we will compare the cost of computations using Newton's method with implicit vs.\ explicit Coriolis integration for various values of the Ekman number by comparing the number of GMRES iterations necessary for~each.

\subsection{Traveling~Waves}
\label{sec:traveling_waves}

Newton's method can also be used to compute traveling waves in the same way. Azimuthal rotating waves satisfy $\bU(\phi,t) = \tilde{\bU}(\phi-Ct)=\tilde{\bU}(\tilde{\phi})$,
where $C$ is an unknown wavespeed.
Thus, $\pd_t \bU=-C\pd_{\tilde{\phi}}\tilde{\bU}$. Substituting into
Equation \eqref{eq:ode} and dropping the tildes leads to
\begin{align}
0 &=  \mN(\bU)  +C \pd_\phi \bU + \mL \bU 
\label{eq:twdefine}
\end{align}
{The explicit} 
 portion of the timestep is augmented to include $C\pd_\phi$ as well as $\mN$:
\begin{align}
\bU(t+\dt) &= (I - \dt \mL)^{-1}\left(I + \dt(\mN + C \pd_\phi )\right)\bU(t)
\end{align}
{If $\bU$} is expressed in terms of spherical harmonics in which the $\phi$ dependence is trigonometric, the~action of $C\pd_\phi$ on a Fourier component $\bU_m$ is merely $C\pd_\phi \bU_m = Cim\bU_m$. 

To use Newton's method to determine the unknown field $\bU$ and wavespeed $C$, 
we substitute $\bU \rightarrow \bU-\bu $ and $C \rightarrow C-c$ into Equation \eqref{eq:twdefine}:
\begin{align}
0 &=  \mN(\bU-\bu) + (C-c)\pd_\phi (\bU-\bu)+ \mL (\bU-\bu) 
\end{align}
{Expanding} around $(\bU,C)$ and truncating at first order leads to the linear system that must be solved for the decrements $(\bu,c)$:
\begin{align}
(\mN_\bU +  C\pd_\phi +\mL) \bu  + c \pd_\phi \bU  = (\mN + C\pd_\phi + \mL) (\bU)
\label{eq:twNewton}
\end{align}
%
{The preconditioner} remains $\mP(\dt)=(I-\dt\mL)^{-1}\dt$ with large $\dt$, since $\mL$ continues to be responsible for the large condition number of Equation \eqref{eq:twNewton}.
\begin{align}
\mP(\dt) \left[\left(\mN_\bU +  C\pd_\phi +\mL\right)\bu + c \: \pd_\phi \bU \right ]= \mP (\dt)(\mN + C\pd_\phi + \mL) (\bU).
\end{align}
{An additional} equation must be added to the system to compensate for the additional unknown. We choose a phase condition, more specifically we require the imaginary part of a single component (temperature, toroidal or poloidal field; radial and angular and azimuthal mode) of $\bu$  be zero, i.e.,~that the corresponding {\it {phase component} 
} (whose index we shall call $J$) of $\bU$ remains~unchanged.

This simple choice suggests a trick for retaining the size of the unknown $\bu$ rather than using augmented fields, since $\bu_J$ is no longer an unknown. The~routine which acts on $(\bu,c)$ is defined such that $c$ is stored in $\bu_J$. At~the beginning of an action, $c$ is extracted, $\bu_J$ is set to zero, and~the explicit part of the action on $\bu$ is carried out. Then, we multiply the stored value of $\pd_\phi \bU$ by $c$ and add the result to the explicit part of the action. When the Krylov method converges and returns $\bu$, we must again extract $c$ from $\bu_J$, after~which $\bu_J$ is set to zero. Effectively, although~$C$ and $c$ have been added to the unknowns for the Newton method and for the linear equation, $\bU_J$ and $\bu_J$ are no longer~unknowns.

\subsection{Continuation}
\label{sec:continuation}

We compute solutions along a branch via Newton's method as described above. We do not impose additional equations, such as requiring a new solution along the branch to be a certain distance from a prior solution or that the increment along the branch must be perpendicular to some direction. Therefore, the only additional ingredients that must be introduced to discuss continuation are the choice of initial estimate for each solution along the branch and the parameters that are prescribed for the Newton~iteration.

If the continuation is in the Rayleigh number, we increment or decrement this parameter according to the number of Newton iterations required for convergence in the previous step as follows~\cite{boronska2010extreme_II}:
\begin{equation}\label{eq:new_Ra}
    Ra^{(i+1)} = Ra^{(i)} + \Delta Ra = Ra^{(i)} + \alpha (Ra^{(i)} - Ra^{(i-1)})
\end{equation}
\noindent where $\Ra^{(i)}$, $\Ra^{(i-1)}$ denote the Rayleigh numbers used in the two previous continuation steps, $\Delta Ra$ is the increment or decrement, and~$\alpha$ is defined by
\begin{equation}\label{eq:alpha_cont}
    \alpha = \frac{N^{\rm opt} + 1}{N^{(i)} + 1}
\end{equation}
$N^{\rm opt}$ is the target number of Newton iterations. If~$N^{(i)}=N^{\rm opt}$, then $\Delta Ra$ remains unchanged, whereas $\Delta Ra$ is reduced (increased) if $N^{(i)}>(<) N^{\rm opt}$. The~choice of $N^{\rm opt}$ is guided by two considerations. The~first is the level of sampling desired along the branch. The~second is economy: a smaller value of $\alpha$ will engender more values of $\Ra$, but~each calculation will be faster. For~our computations, we fixed $N^{\rm opt}$ to be between 4 and~6. 

To choose an initial estimate $\bU$ for the next solution along a branch at $\Ra^{(i+1)}$, first-order and even zero-th order extrapolation (i.e.,\ using just the previously computed field) can be used to follow a smooth and monotonically varying branch. However, quadratic extrapolation is essential for going around turning points. Lagrange interpolation uses the three previous Rayleigh numbers $\Ra^{(i)}$, $\Ra^{(i-1)}$, $\Ra^{(i-2)}$ and the new Rayleigh number $\Ra^{(i+1)}$ from Equation \eqref{eq:new_Ra} to determine coefficients $a,b,c$ such that
\begin{subequations}
\label{eq:lagrange}
\begin{align}
Ra^{(i+1)} &= a Ra^{(i)} + b Ra^{(i-1)} + c Ra^{(i-2)} .\label{eq:lagrangeRa}
\intertext{{We} then use $a,b,c$ to form a new estimate of the solution by setting}
    \bU^{(i+1)} &= a \bU^{(i)} + b \bU^{(i-1)} + c \bU^{(i-2)}\label{eq:lagrangeU}
\end{align}
\end{subequations}
{Newton} iterations are then used to refine $\bU$ until the norm of Equation \eqref{eq:twdefine} is $10^{-7}$. This procedure requires saving the three previous solution vectors and Rayleigh~numbers.

\subsection{Turning~Points}
We now turn to the more complicated matter of extrapolating near saddle--node bifurcations, at~which $\bU$ ceases to be a single-valued function of $\Ra$. 
The code detects that the current step is in the vicinity of a turning point by comparing the relative changes between the solution vector $\bU$, and~the Rayleigh number, i.e.,\ we compare
\begin{equation} \label{eq:tp_detection_Ra}
\begin{split}
        \left| \frac{\Delta U_J}{U_J} \right| \equiv \left| \frac{U_J^{(i)} - U_J^{(i-1)}}{U_J^{(i)}} \right| \qquad {\rm with} \qquad
       \gamma \left| \frac{\Delta Ra_{~}}{Ra_{~}} \right| \equiv \gamma \left| \frac{Ra^{(i)_{~}} - Ra^{(i-1)}}{Ra^{(i)}_{~}} \right|
\end{split}
\end{equation}
\noindent where $J$ is the index denoting the element of the solution vector $U^{(i)}$ of the highest absolute value and $\gamma$ is a constant factor that makes the two magnitudes comparable. For~our computations, we fixed this constant to be between 10 and~100.

Since the normal form of a saddle--node bifurcation at $(\mu,x)=(0,0)$ is $\dot{x} = \mu - x^2$, dependent variables ($x$, $\bU$) vary like the square root of the control parameter ($\mu$, $\Ra-Ra_c$) near a turning point. Consequently, as~the current solution approaches a turning point, the~relative difference $|\Delta U_J/U_J|$ increases until it is greater than  $\gamma|\Delta Ra/Ra|$. It is at this stage that the code switches to extrapolating quadratically, using the component $U_J$ as the independent variable instead of the Rayleigh number. This allows $\Delta Ra$ to change sign and the continuation to go around the turning point. 
We replace Equations \eqref{eq:new_Ra} and \eqref{eq:lagrangeRa} by analogous equations in $\bU_J$ to determine $\bU_J^{(i+1)}$ and $a,b,c$, and~then 
replace 
\mbox{Equation \eqref{eq:lagrange}~by}
\begin{subequations}
\begin{align} \label{eq:quad_extra_Ra}
    U_j^{(i+1)} &= a U_j^{(i)} + b U_j^{(i-1)} + c U_j^{(i-2)} \mbox{  for }j\neq J \mbox{,  and~ }  \\
    Ra^{(i+1)} &= a Ra^{(i)} + b Ra^{(i-1)} + c Ra^{(i-2)}
\end{align}
\end{subequations}
to set the estimate of the new solution. Note that we have only changed the method of determining the initial guess. $\Ra$ remains fixed to $\Ra^{(i+1)}$ during the Newton step, while all the elements of $\bU$ are allowed to~vary.

Continuation follows in this manner until eventually $|\Delta \bU_J/\bU_J|$ exceeds $\gamma|\Delta Ra/Ra|$, at~which point the code switches back to using the Rayleigh number as the independent variable. The~parameter $\gamma$ can be obtained by analyzing the behavior of the continuation near a turning point. Indeed, if~$\gamma$ is too large, then the code will continue to use $\Ra$ as the independent variable past the point at which there is no solution for the next continuation step, and~Newton's method will not converge. Should this situation arise, the~continuation can be restarted at the last converged solution using a suitably reduced value of $\gamma$.
{ Although the parameter $\gamma$ is necessary for comparing $|\Delta U_J/U_J|$ and $|\Delta Ra/Ra|$ and thus for determining when to switch continuation variables, its value has no influence on the location of the turning point or of the solution branch. The~choice of $\gamma$ only affects the speed of continuation, as~we determined by varying its value near a turning point. See also~\cite{boronska2010extreme_II}.}

\section{Branch~Following}
\label{sec:branch_following}

\subsection{Continuation in Rayleigh~Number}

{ Our first computations with the implicit Coriolis method validated the new code by comparing the branches it produced for $\Ek=10^{-3}$ with those in~\cite{feudel2013multistability}, which were obtained using an entirely different continuation code. These branches were presented in Figure~\ref{fig:Feudel} above.}
As explained in Section~\ref{sec:numerical_methods}, to~search for solutions having an $M$ azimuthal periodicity, we perform our computations in a fraction of the annulus of azimuthal width $2 \pi / M$. The~solutions are computed independently of their stability; most of the solutions we compute are unstable, either within the restricted domain $0\leq \phi < 2\pi/M$, or~to instabilities which break this restriction such as rotating waves with a different $M$. 

\begin{figure}
\subfloat[\centering]
{\includegraphics[height=6cm]
{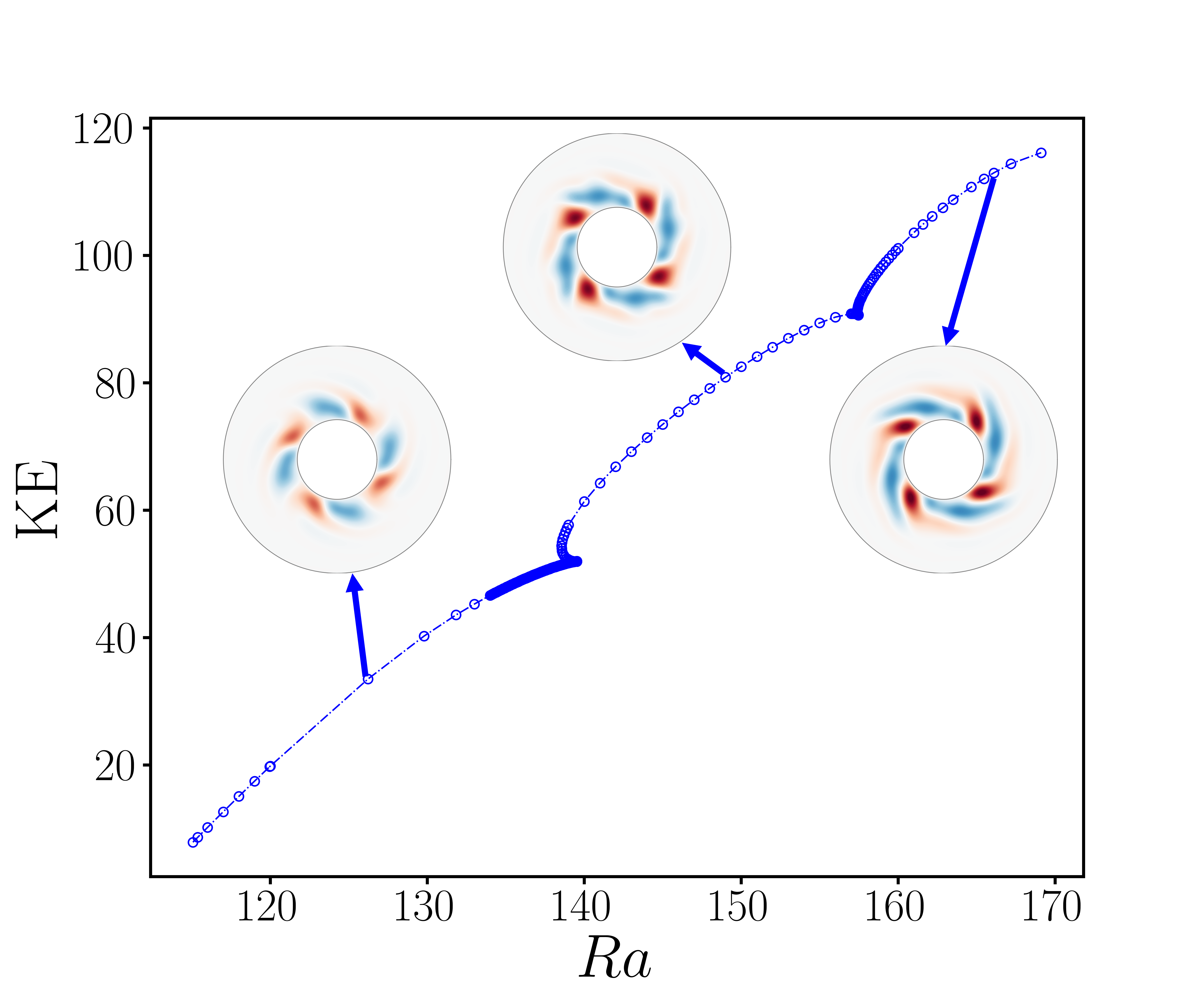}}
\subfloat[\centering]
{\includegraphics[height=6cm]
{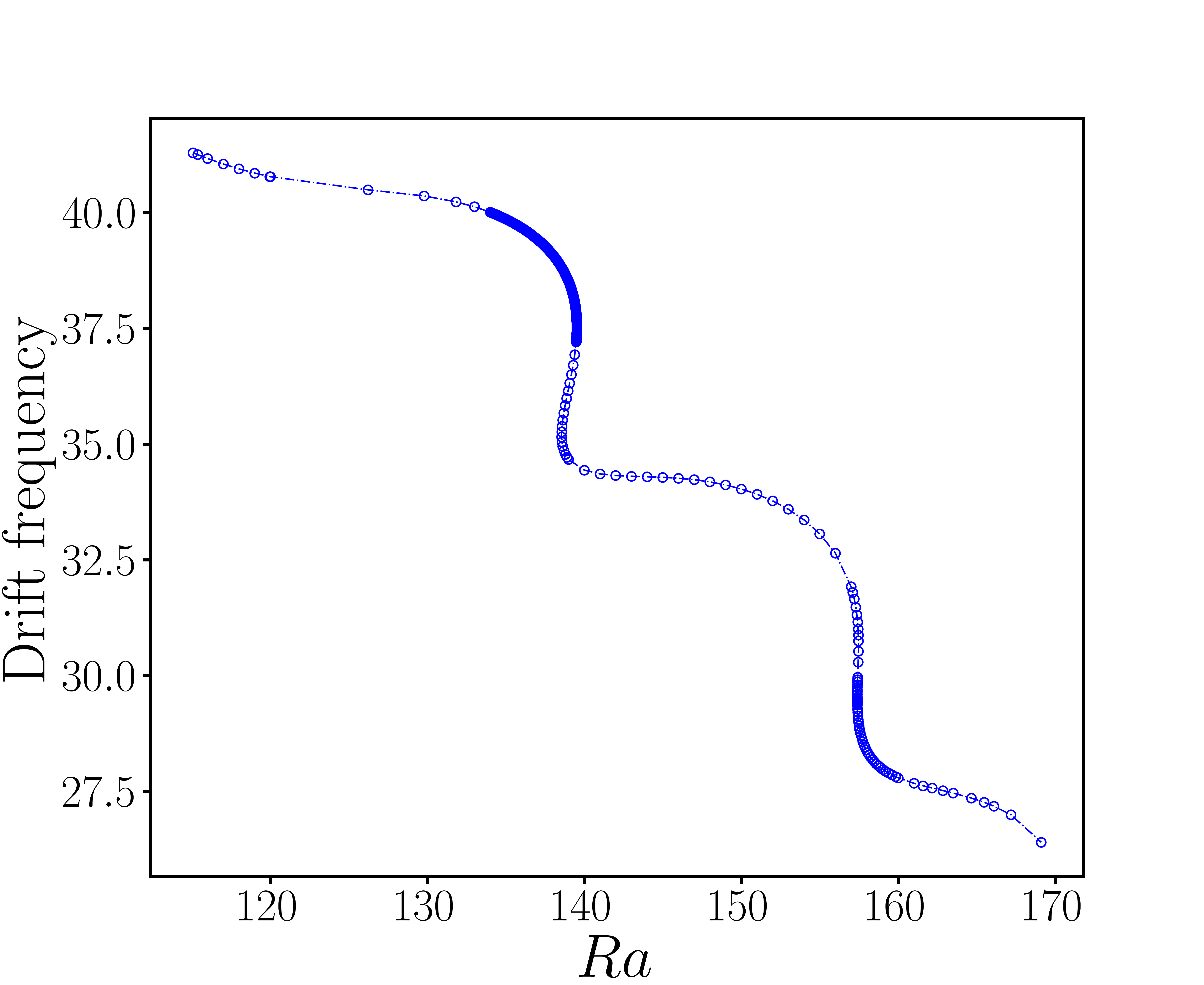}}
\caption{{Bifurcation} 
 diagram of rotating wave RW$_4$ for $\Ek=3.53\times 10^{-5}$ as a function of $\Ra$. (\textbf{a}) Kinetic energy. (\textbf{b}) Drift frequency. The~branch contains three long smooth regions of almost constant drift frequency separated by two short intervals (at $\Ra\approx 140$ and $\Ra\approx 158$) of rapid change containing saddle--node bifurcations. The~resolution used for this computation was $(N_r, N_\theta, N_\phi \times M) = (60, 80, 40 \times 4)$.} \label{fig:RW4_Ek_3p53_m5}
\end{figure}

We then used the continuation code to compute branches of rotating waves at lower values of the Ekman number. 
Fixing the Ekman number, we carried out continuation in Rayleigh number $\Ra$. Most of the branches that we computed showed no unusual features, varying smoothly and monotonically down to their threshold at a supercritical Hopf bifurcation. However, a~few branches presented some interesting non-monotonic behavior, which we show in order to display the capacities of our code. Figure~\ref{fig:RW4_Ek_3p53_m5} tracks the Rayleigh-number dependence of the RW$_4$ branch for $\Ek=3.53\times 10^{-5}$. This branch consists of three smooth regions that are separated from one another by short intervals which each contain two saddle--node bifurcations and rapid changes in drift~frequency. 

\begin{figure}
\subfloat[\centering]
{\includegraphics[height=6cm]{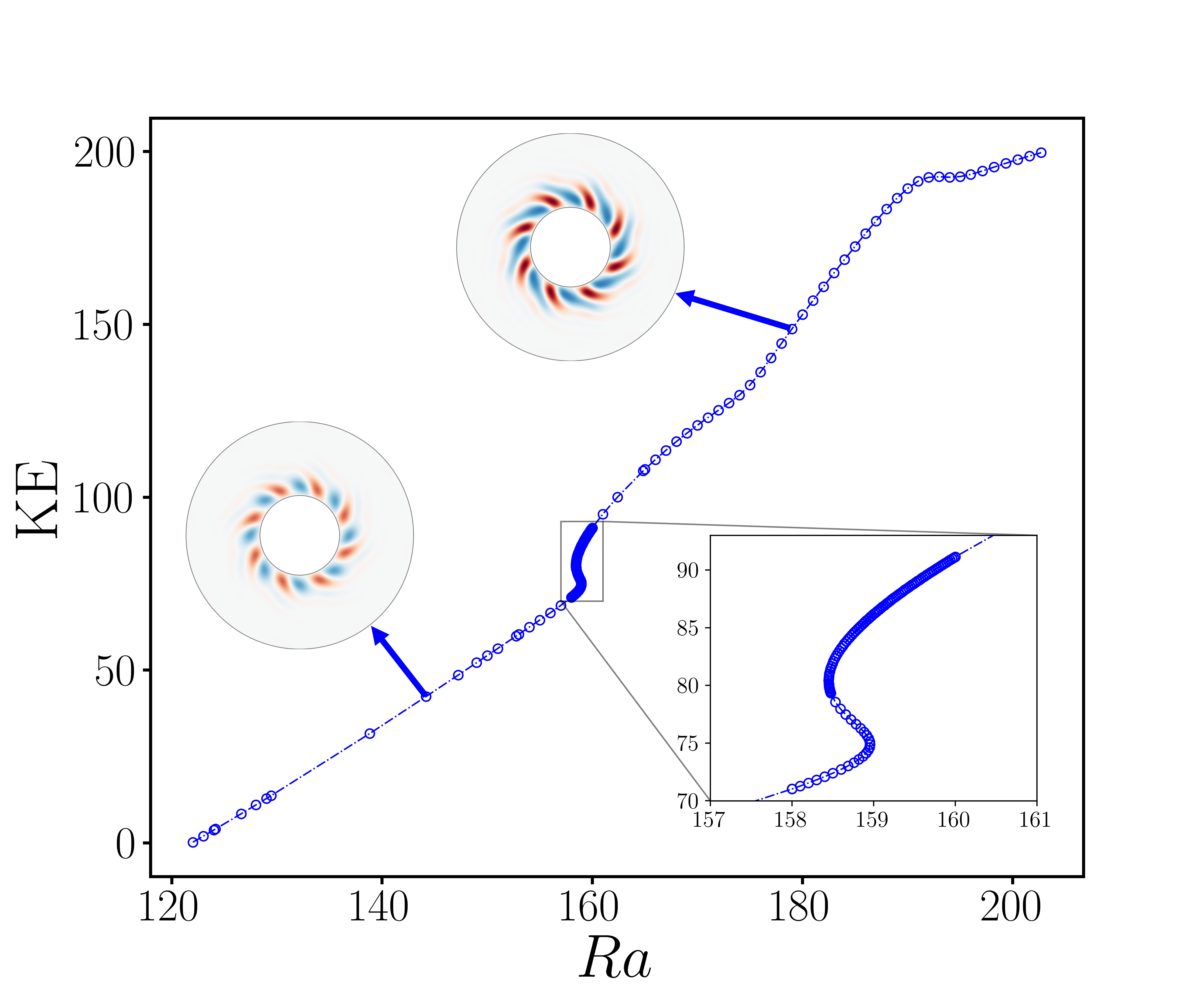}}
\subfloat[\centering]
{\includegraphics[height=6cm]
{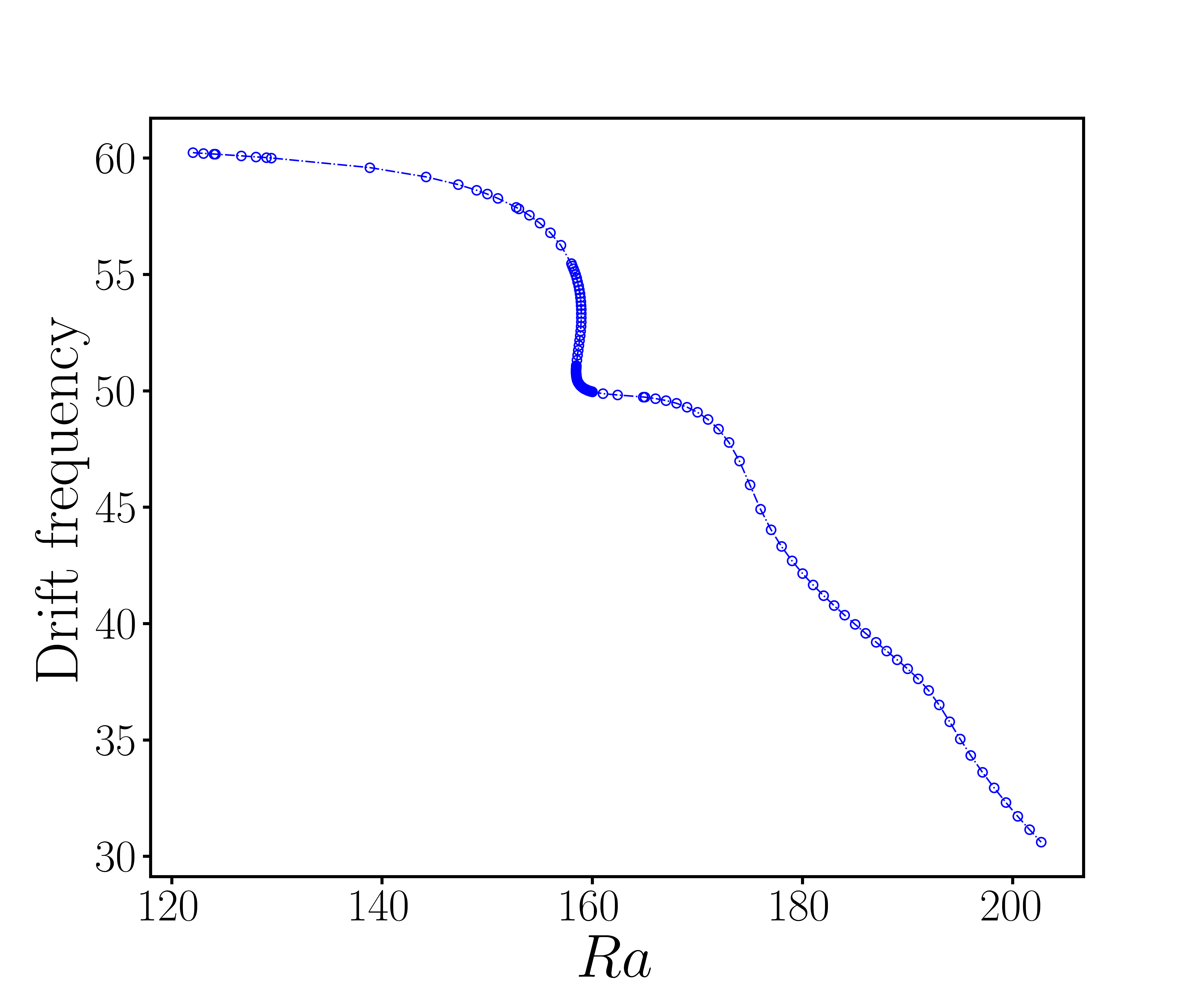}}
\caption{{Bifurcation}  
 diagram of rotating wave RW$_8$ for $\Ek=1.26\times 10^{-5}$ as a function of $\Ra$. (\textbf{a})~Kinetic energy. (\textbf{b}) Drift frequency. The~branch contains a single interval of rapid change at $\Ra\approx 159$. The~resolution used for this computation was $(N_r, N_\theta, N_\phi \times M) = (68, 100, 28 \times 8)$.} \label{fig:RW8_Ek_3p16_m5}
\end{figure}

We also computed an RW$_8$ branch for $\Ek=1.26\times 10^{-5}$, as shown in Figure~\ref{fig:RW8_Ek_3p16_m5}. Indeed, for~this value of $\Ek$, rotating waves with an azimuthal wavenumber of 8 are more appropriate, i.e.,\ more likely to be stable.  This branch also contains a plateau in drift frequency adjacent to a short interval at $\Ra\approx 159$ of rapid change delimited by two saddle--node bifurcations. { These rapid rises and plateaus in drift frequency should have a physical or at least a phenomenological explanation, but~we have not yet been able to find one.}

%
%

\subsection{Continuation in Ekman Number and in~Resolution}

We adapted our method to follow branches in Ekman number, varying $\Ek$ on a logarithmic scale. We replace $\Ra$ with $\log_{10}(\Ek)$ in Equation
\eqref{eq:new_Ra} as follows:
\begin{align}\label{eq:new_EK}
    \log_{10} (\Ek^{(i+1)}) &=  \log_{10} (\Ek^{(i)}) + \Delta [\log_{10}(\Ek)] \nonumber\\&= \log_{10}(\Ek^{(i)}) + \alpha (\log_{10}(\Ek^{(i)}) -\log_{10}(\Ek^{(i-1)})) 
\end{align}
where the computation of $\alpha$ remains as in Equation \eqref{eq:alpha_cont}. In~this case, we used $N^{\rm opt} = 3$ for all of the simulations. Furthermore, in~order to go around turning points, we now compare $|\Delta \bU_j/\bU_j|$ and $\gamma |\Delta [\log_{10}(\Ek)]/\log_{10}(\Ek)|$, where we chose $\gamma$ between 400 and~500. 

Apart from these minor changes, there are several major differences between performing continuation in Ekman and Rayleigh numbers.
The first is that the matrices that represent the left-hand-sides of Equations \eqref{eq:expcor_e} and \eqref{eq:expcor_f} or Equations 
\eqref{eq:impcor_e_final} and \eqref{eq:impcor_f_final}
must be recomputed every time a new value of $\Ek$ is chosen. Indeed, while the Rayleigh number only appears in the right-hand side of these equations,  $\Ek$, is also present in the diffusion terms, which are treated implicitly regardless of 
how the Coriolis term is handled.
This means that extra work must be performed 
at the beginning of every continuation~step.

Secondly, as~the Ekman number is decreased, the~critical Rayleigh number for the onset of convection increases. That is, rotation stabilizes the configuration against convection. A~classic result~\cite{busse1970thermal} is
that the critical value of the usual thermal Rayleigh number $\Ra_{\rm therm}$ varies like $\Ek^{-4/3}$, so that the critical value of our rotational Rayleigh number $\Ra_{\rm rot}=\Ra_{\rm therm} \Ek/\Pra$ varies like $\Ek^{-1/3}/\Pra$. Therefore, instead of keeping $\Ra$ fixed, we have increased $\Ra$ every time we decrease $\Ek$, so as to remain at approximately the same distance from the convective threshold. 

Third, as~the simulation ventures towards lower Ekman numbers, the~fields require a higher resolution. This is clearly seen from the visualizations in Figure~\ref{fig:four_rotating_waves}, which show that the radial extent and azimuthal wavelength decrease with decreasing $\Ek$. To~achieve this, grid refinement is introduced in the code such that the spectral resolution is increased by 20\% whenever under-resolution is detected in the Chebyshev or the spherical harmonic modes. This is performed by introducing a threshold for the amplitude ratio between the mode of highest absolute value and of highest wavenumber. When this ratio exceeds the threshold, the~code calls a grid refinement subroutine which first deallocates the previous grid, then creates a new one using the new resolution and finally represents the fields from the last continuation step on this new grid by filling in the new modes with~zero.

{ This procedure, called Fourier
  interpolation~\cite{lanczos1938trigonometric,trefethen2000spectral},
  involves no loss of accuracy or stability. The~most recent Newton
  step is then recomputed, using the representation with the finer
  resolution as an initial condition, and~the path following continues
  as intended. Most of our runs were carried out with an
  under-resolution threshold (ratio of the highest-amplitude mode to
  that of the highest wavenumber) of $10^{-6}$. Decreasing this
  threshold to $10^{-7}$ led to minimal gains in accuracy at a high
  cost in CPU time.}

\begin{figure}
\subfloat[\centering]
{\includegraphics[height=6cm]{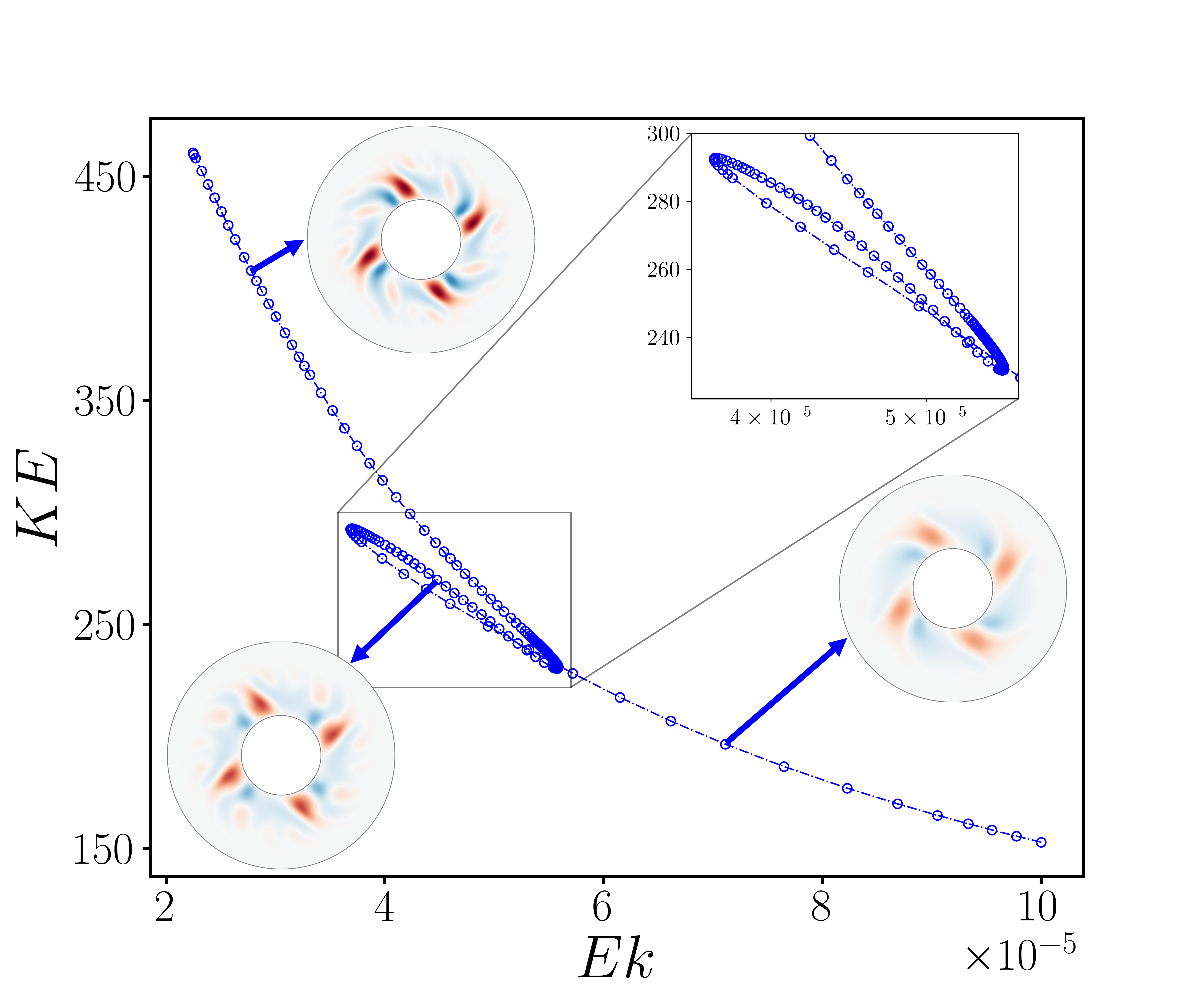}}
\subfloat[\centering]
{\includegraphics[height=6cm]
{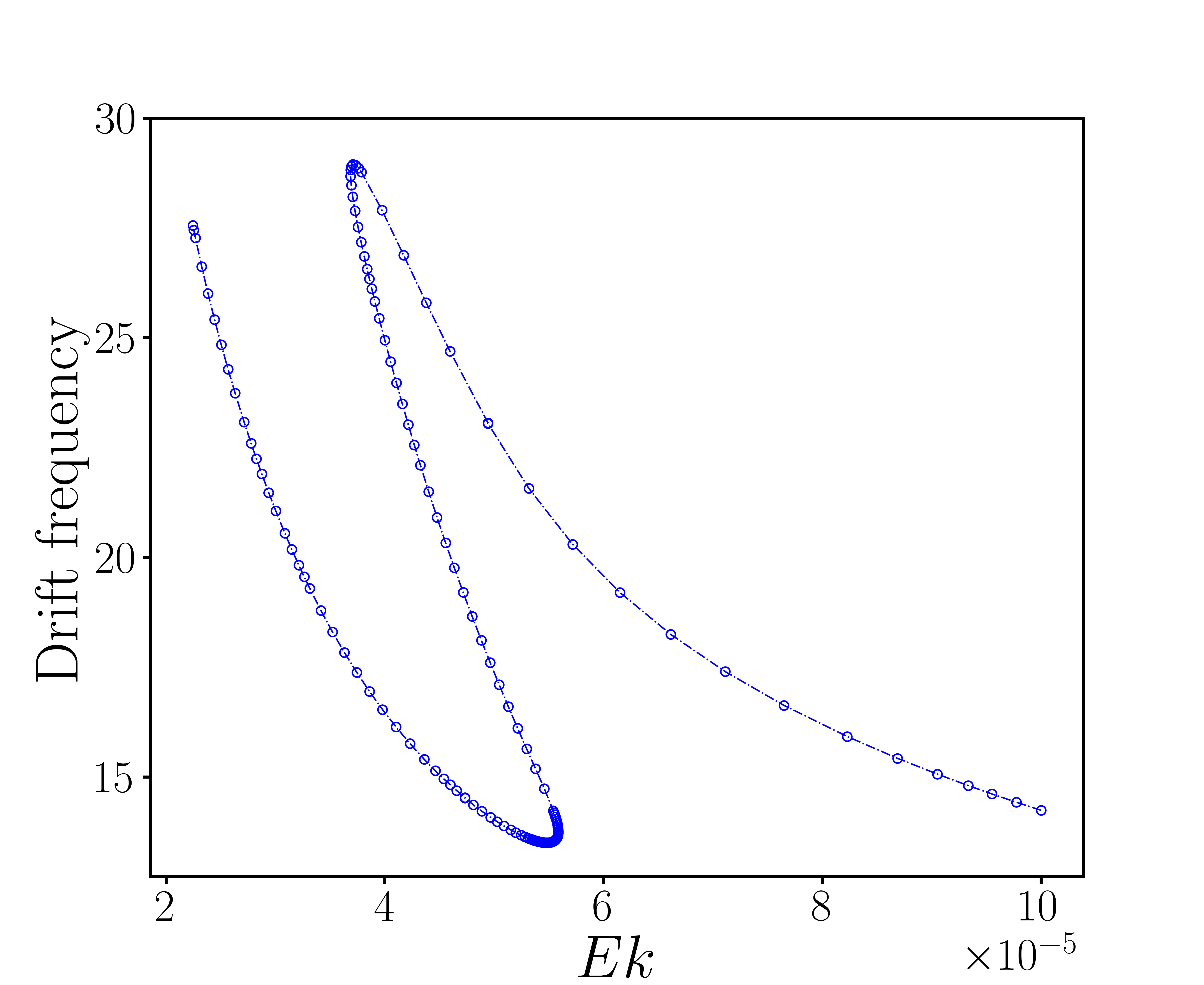}}
\caption{Bifurcation 
 diagram of the rotating wave RW$_4$ as a function of $\Ek$, while fixing \mbox{$\Ra=6.5\times \Ek^{-1/3}$}. (\textbf{a}) Kinetic energy. (\textbf{b}) Drift frequency. Three solution branches exist over the range $4\times 10^{-5} \lesssim \Ek \lesssim 6 \times 10^{-5}$. The~resolutions used for this computation ranged from $(N_r, N_\theta, N_\phi \times M) = (68, 92, 48 \times 4)$ to $(80, 136, 70 \times 4)$.}  \label{fig:RW4_Ek_cont}
\end{figure}

Figure~\ref{fig:RW4_Ek_cont} shows the result of continuation in $\Ek$. (Recall that $\Ra$ is kept proportional to $\Ek^{-1/3}$.) This branch showed three coexisting solutions over the range $4\times 10^{-5} \lesssim \Ek \lesssim 6\times 10^{-5}$. Our grid refinement algorithm increased the resolution by about 20\% in $r$ and by 50\% in $\theta$ and $\phi$. We emphasize that most continuations showed monotonic behavior and we have deliberately chosen to present those that did not.

\section{Timing~Comparisons}
\label{sec:timing_comparisons}

\begin{figure}
\centering
\includegraphics[width=0.9\columnwidth]{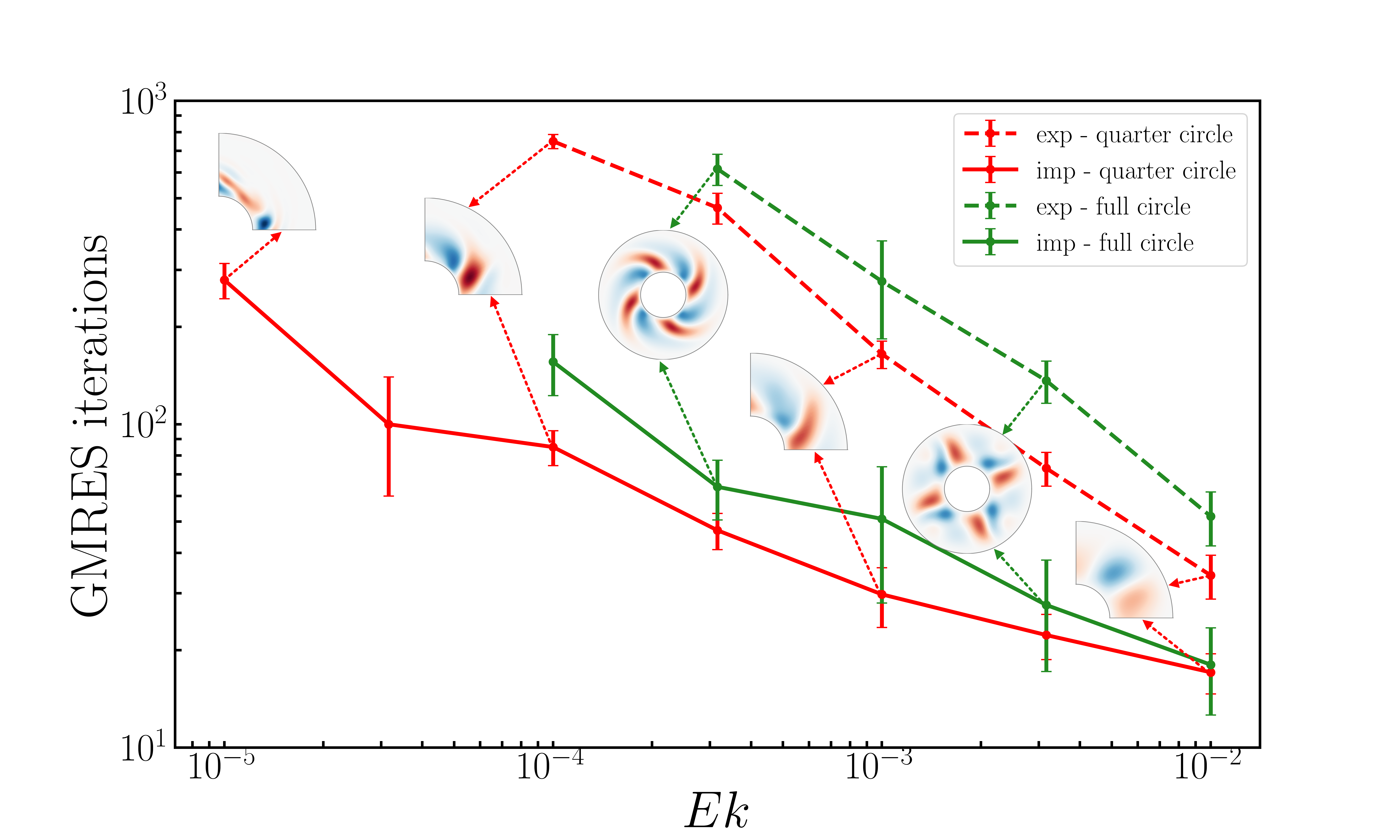}
\caption{Total number of matrix--vector actions required by nested Newton-GMRES algorithm to compute RW$_4$ as a function of $\Ek$ with explicit (dashed) and implicit (solid) implementation of Coriolis force. An~average is taken over a branch of $\Ra$ values. The~number of actions required by the explicit algorithm is always greater than that required by the implicit algorithm, with~the ratio between them increasing from approximately 2 at $\Ek=10^{-2}$ to approximately 9 at $\Ek=10^{-4}$. Using the full domain $0\leq\phi < 2\pi$ (green) instead of the reduced domain $0\leq\phi < \pi/2$ (red) increases the number of GMRES iterations, as~well as the cost of each iteration. On~average, the~full domain requires 1.5~times more actions than the reduced domain counterpart. The~explicit algorithm is prohibitively time-consuming for $\Ek< 10^{-4}$. The~resolutions used in this case were $(N_r, N_\theta, N_\phi \times M) = (46, 72, 128 \times 1)$ for the computations in the full circle and from $(46, 72, 32 \times 4)$ to $(76, 80, 40 \times 4)$ for those in the quarter~circle.}
\label{fig:RW4_quarter}
\end{figure}

\begin{figure}
\centering
\includegraphics[width=0.9\columnwidth]{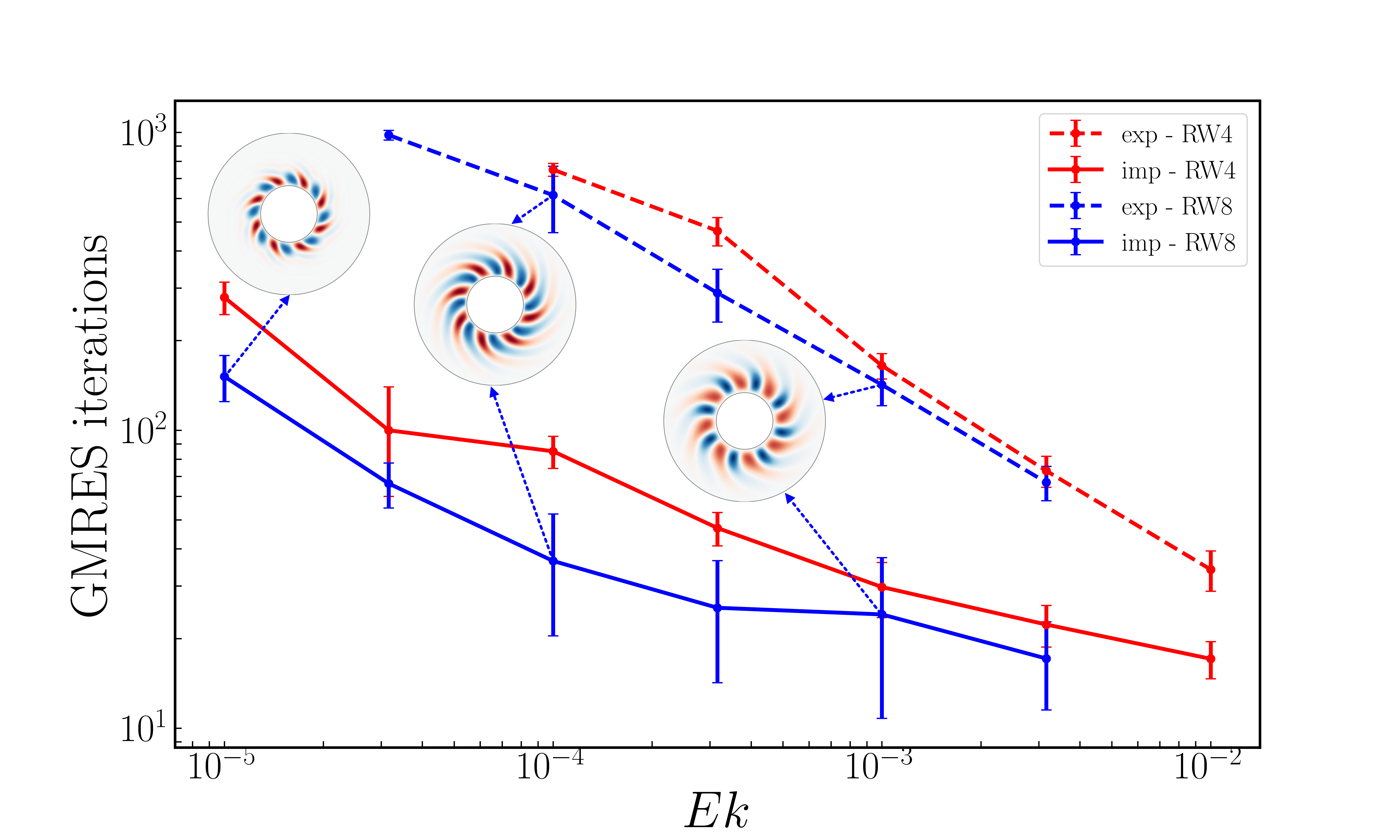}
\caption{Total number of matrix--vector actions required by the nested Newton-GMRES algorithm to compute RW$_4$ (red) and RW$_{8}$ (blue) as a function of $\Ek$ with an explicit (dashed) and implicit (solid) implementation of Coriolis force. For clarity, the~insets show the full domain, although~the computations are carried out in reduced domains $0\leq\phi < 2\pi/M$. An~average is taken over many values of $\Ra$. The number of actions required by the explicit algorithm is always greater than that required by the implicit algorithm, with the ratio between them increasing from approximately 4 at $\Ek=3 \times 10^{-3}$ to 15 at $\Ek=3\times 10^{-5}$, below~which the explicit algorithm is prohibitively time-consuming. The~resolutions used for the RW$_8$ computations ranged from $(N_r, N_\theta, N_\phi \times M) = (64, 100, 28 \times 8)$ to $(76, 124, 32 \times 8)$.}
\label{fig:RW8_gmres_its}
\end{figure}

\begin{figure}[H]
\centering
\includegraphics[width=0.9\columnwidth]{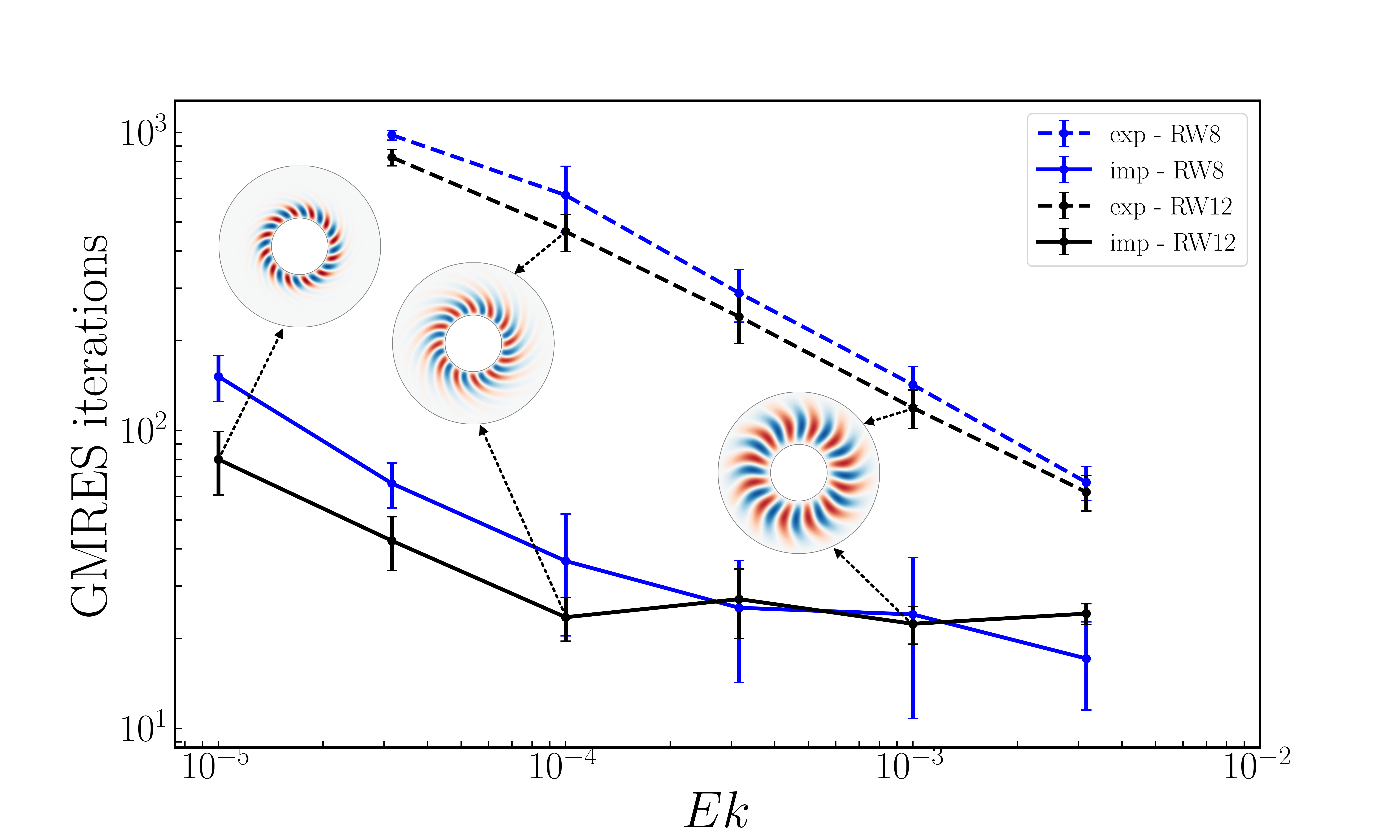}
\caption{Total number of matrix--vector actions required by the nested Newton-GMRES algorithm to compute RW$_{12}$ (black) and RW$_8$ (blue) as a function of $\Ek$ with explicit (dashed) and implicit (solid) implementation of Coriolis force. For~clarity, the~insets show the full domain, although~the computations are carried out in reduced domains $0\leq\phi < 2\pi/M$. An~average is taken over a branch of $\Ra$ values. The number of actions required by the explicit algorithm is always greater than that required by the implicit algorithm, with the ratio between them increasing from approximately 2.5 for $\Ek=3 \times 10^{-3}$ to 20 for $\Ek=3\times 10^{-5}$, below~which the explicit algorithm fails. The~resolutions used for the RW$_8$ computations ranged from $(N_r, N_\theta, N_\phi \times M) = (60, 112, 20 \times 12)$ to $(90, 184, 32 \times 12)$.}
\label{fig:RW12_gmres_its}
\end{figure}

Newton's method is very fast, typically requiring 3-10 iterations. The~bottleneck in applying Newton's method to systems of partial differential equations is always the solution of the linear systems. In~our case, this is quantified by the total number of GMRES iterations necessary to compute a new state along the branch. We find that the number of iterations depends on the Ekman number, but~is fairly independent of the Rayleigh number. We therefore average the number of GMRES iterations over all the points computed in a branch for a fixed Ekman number. We do this for the Newton method with both implicit and explicit Coriolis to produce the curves in Figures~\ref{fig:RW4_quarter}--\ref{fig:RW12_gmres_its} for RW$_4$, RW$_8$, and~RW$_{12}$. The~ratio of the number of GMRES iterations between the explicit and implicit methods ranges from 2, for~RW$_4$ at $\Ek=3\times 10^{-2}$, to 20 for RW$_{12}$ at $\Ek=3\times 10^{-5}$. This ratio would surely continue to increase as $\Ek$ decreases, but~the explicit Coriolis calculation became  prohibitively expensive below $\Ek=3\times 10^{-5}$. Figure~\ref{fig:RW4_quarter} also demonstrates that carrying out the calculation in the restricted domain $0\leq \phi\leq 2\pi/M$ requires fewer GMRES iterations (in addition to each iteration being at least $M$ times faster).

\begin{figure}[H]
\centering{
\includegraphics[width=0.8\columnwidth]{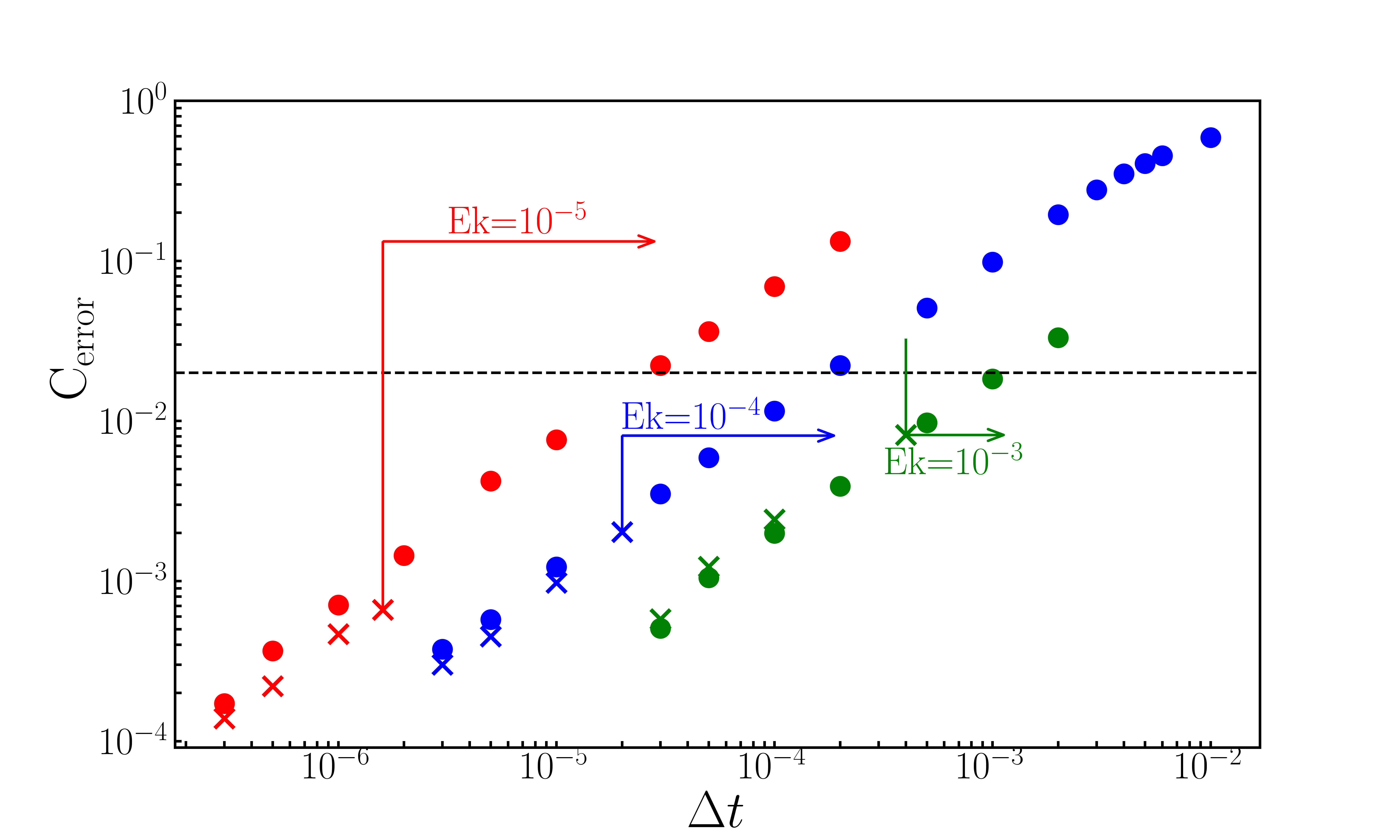}}
 \caption{Accuracy of the time-dependent calculation of rotating waves RW$_M$ as a function of timestep $\dt$ using implicit (circles) and explicit (crosses) timestepping of the Coriolis term. Shown here is the relative error $C_{\rm error}\equiv|(C_{\dt}-C_{\rm exact})/C_{\rm exact}|$ of the drift velocity as a function of the timestep for various sets of parameter values (see table) where $C_{\rm exact}$ is calculated via Newton's method. Both the explicit and implicit methods are second order in time, with~$C_{\rm error}\sim \dt^2$. 
 Vertical lines indicate the limiting $\dt$ above which explicit timestepping diverges and cannot be used. 
 The horizontal dashed line indicates $C_{\rm error}=2\%$.
 The arrows designate the range of $\dt$ 
 for which implicit timestepping is advantageous: it does not diverge and
 $C_{\rm error}$ remains less than $2\%$.  
 { An approximate fit yields the range $0.5\times \Ek^{1.1} \lesssim \dt \lesssim 0.8\times Ek^{0.9}$.}
 See Table~\ref{tab:timestep}.}
 \label{fig:timetable_all}
 \end{figure}

Measuring the economy realized in timestepping is more problematic. Because~explicit integration effectively approximates a decaying exponential by a polynomial, it displays artificial temporal divergence, i.e.,\ numerical instability, if~the timestep is too large. Because~implicit integration instead approximates the exponential by a decaying rational function, the~timestep is not constrained by stability. 
The timestep is, however, still constrained by accuracy. A~particularly demanding criterion for judging accuracy is the wavespeed~\cite{marcus1984simulation}. 
We compute the relative errors in wavespeeds $C_{\rm error}\equiv(C_{\dt}-C_{\rm exact})/C_{\rm exact}$, where $C_{\rm exact}$ is obtained from Newton's method and hence has no timestepping error. 
Figure~\ref{fig:timetable_all} presents the relative errors as a function of $\Ek$ and $\dt$, with~the other parameters set to the values below the figure. $C_{\rm error}$ obtained from the explicit and implicit methods are the same or very close for the same value of $\dt$. 
$C_{\rm error}$ increases like $(\dt)^2$, indicating that both time-integration methods are second-order in time. Indeed, for~timestepping, we have not used the first-order scheme given in Equation \eqref{eq:discrete} and used for our Newton algorithm, but~a second-order method combining Runge--Kutta and a predictor--corrector scheme, as~described in~\cite{hollerbach2000spectral}.

For each $\Ek$, there is a minimum $\dt$ above which the explicit Coriolis simulation diverges in time. This value of $\dt$ is indicated in Figure~\ref{fig:timetable_all} for each Ekman number as the left endpoint of an arrow and also in Table~\ref{tab:timestep}. Above~this value of $\dt$, implicit Coriolis timestepping must be used. But~eventually, $\dt$ is so large that the 
results of the implicit method are too inaccurate to be useful.
As an upper limit for the implicit method, we make the arbitrary choice that the relative error on $C_{\rm error}$ must be less than 2\%.
These limiting values of large $\dt$ are indicated as the right endpoints of the arrows in Figure~\ref{fig:timetable_all} and in Table~\ref{tab:timestep}. Hence, for~each $\Ek$, there is a large range of $\dt$-values that can be used only for implicit timestepping and whose error does not exceed $2\%$. 
For example, for~$\Ek=10^{-5}$, the~allowable timestep (<\:2\% error) for the implicit Coriolis method is almost 20 times that at which the explicit method diverges. These ranges are given in Table~\ref{tab:timestep}.
{Fitting to the two smaller values of $\Ek$, both endpoints of the range are approximately proportional to $\Ek$: the explicit method diverges for $\dt\gtrsim 0.5 \;\Ek^{1.1}$ while the error for the implicit method remains less than 2\% for $\dt \lesssim 0.8\;  \Ek^{0.9}$.}
{ If a 2\% error in the wavespeed is considered to be too large, the~data in Figure~\ref{fig:timetable_all} can be used to determine the largest $\dt$ that achieves the desired error, by~drawing a horizontal line at the desired level.}
Although these results are specific to the parameter values and flows that we have simulated, they imply that implicit Coriolis integration can be orders of magnitude faster as $\Ek$ decreases, enabling the simulation of convection in a spherical annulus at low Ekman~numbers.


\begin{table}[H]
\caption{Parameters 
 used for testing and useful range of $\dt$ for implicit~timestepping.}
\centering{
\begin{tabular}{ccccc|ccc}
\toprule
&&&&&\multicolumn{2}{c}{\qquad\quad\boldmath$\dt$} &\\
$\Ek$ & $\Ra$ & $M$ & $(N_r, N_\theta, N_\phi \times M)$ & $C_{\rm exact}$ & Explicit Diverges & 2\% Error & Ratio\\
\hline
$10^{-3}$ & $120$ & $4$ & $(46, 72, 32 \times 4)$ & $-2.7647$ & $4.0\times 10^{-4}$ & $1.1\times 10^{-3}$ & 3\\
$10^{-4}$ &  $130$& $8$ & $(60, 80, 40 \times 4)$ & $~~3.9998$ & $2.0\times 10^{-5}$ & $1.9\times 10^{-4}$ & 9\\
$10^{-5}$ & $140$& $12$ & $(76, 136, 24 \times 4)$ & $~~51.2817$ & $1.6\times 10^{-6}$ & $2.8\times 10^{-5}$ & 17\\\hline
\end{tabular}}
\label{tab:timestep}
\end{table}

{ We studied the transition to chaos for $\Ek=5 \times 10^{-6}$. Explicit Coriolis timestepping requires $\Delta t \lesssim 5\times 10^{-7}$. Using the implicit method with $5~\dt$ and then $10~\dt$ allowed us to simulate five or 10 times longer and to tentatively confirm the temporally chaotic behavior we observed with the explicit method. If~we wish to study this phenomenon further, we will decrease $\dt$ in order to obtain more accurate results.}

As discussed in Section~\ref{sec:ImpCor}, treating the Coriolis term implicitly requires the use of block pentadiagonal matrices rather than block diagonal matrices. However, the~additional cost is negligible. Part of the reason for this is that, in pseudospectral methods, the~cost of carrying out the nonlinear terms (or the linearized nonlinear terms), more specifically the transforms to and from spectral space, is by far the most time-consuming portion of the calculation. Thus, the~cost of a timestep is barely increased by the implicit treatment of the Coriolis term. In~timestepping, if~$\dt$ can be increased by a factor of 100 (assuming the accuracy is still acceptable), then the simulation to the same final time costs (almost exactly) 100 times less. In~Newton's method, if~10 times fewer actions of the Jacobian (i.e., linearized timesteps) are required for a Newton step, then the computation will be (almost exactly) 10~times~faster. 

\section{Discussion}
\label{sec:discussion}

\subsection{Differences with Other~Methods}

We now describe the differences between Stokes preconditioning and the method used in~\cite{sanchez2004newton,garcia2008antisymmetric,sanchez2013computation,garcia2016continuation} for convection in rotating spheres, as~well as in~\cite{duguet2008relative,gibson2009equilibrium,zheng2024natural_1,zheng2024natural_2,zheng2025natural_3} for other hydrodynamic problems.
These authors seek steady states or traveling waves as roots of $\bU(t+T)-\bU(t)$, where $\bU(t+T)$ is computed from
$\bU(t)$ by carrying out many timesteps, each with a small $\dt$; we therefore call this the integration method. In~contrast, although \mbox{Equations \eqref{eq:continuous} and \eqref{eq:finalrhs}} 
would seem to imply that we too seek roots of $\bU(t+T)-\bU(t)$, this is not at all the case. In~the Stokes preconditioning method, we compute $\bU(t+T)$ from $\bU(t)$ by carrying out a single very large implicit--explicit Euler timestep Equation~\eqref{eq:discrete}, with~a timestep $\dt=T$ that is so large that the difference $\bU(t+\dt)-\bU(t)$ no longer approximates the time derivative. Via Equation~\eqref{eq:finalrhs}, it turns out that this difference is a preconditioned version of the operator whose roots are sought by our method, namely $\mL+\mN$, preconditioned by $\mL$.
Both methods solve the linear systems that are the core of Newton's method via GMRES, which relies on the repeated actions of the Jacobian. For~the integration method, the~action of the Jacobian consists of integrating the linearized equations via many small timesteps, while for us, it consists of taking a single large linearized timestep.
Here, there is a trade-off. Despite the preconditioning displayed by Equation \eqref{eq:finalrhs}, the~Jacobian resulting from Stokes preconditioning is less well conditioned than that which results from taking many small linearized timesteps, and~so more Jacobian actions are required for GMRES to converge to a solution of our linear system Equation \eqref{eq:precond}. However, each action costs less, since it consists of only a single timestep. 
This trade-off---number of actions vs.\ timesteps per action---can be quantified via the total number of timesteps required to compute a steady state or a traveling wave. For~wall-bounded shear flows in the transitional regime, we have found that the Stokes preconditioning method is faster than the integration method by a factor of 11 for plane Couette flow and by a factor of 35 for pipe flow; see~\cite{tuckerman2018order} for details. We have not carried out a timing comparison of the integration and the Stokes preconditioning method for convection in rotating~spheres.

Although the Stokes preconditioning method is much faster (at least in the cases and the parameter regimes that we have studied), the~integration method has the considerable advantage of being far more general. The~Stokes preconditioning method is only capable of computing steady states or traveling waves, while the integration method can compute periodic orbits of all kinds, including, for~example, modulated rotating waves, standing waves, or~pulsing~states.

 A specific difference between our implementation of implicit Coriolis integration and that of~\cite{sanchez2004newton,garcia2008antisymmetric,garcia2010comparison,sanchez2013computation,garciadormy2015,garcia2016continuation} 
 is that we solve the linear systems arising from the implicit treatment of the diffusive and Coriolis terms directly via block pentadiagonal LU decomposition and backsolving, whereas~\cite{sanchez2004newton,garcia2008antisymmetric,garcia2010comparison,sanchez2013computation,garciadormy2015,garcia2016continuation} use Krylov methods to solve these systems.  Time-integration using the explicit, implicit, and semi-implicit treatment of the Coriolis term is compared by~\cite{garcia2010comparison}. Our strategy is to solve linear systems directly as much as possible and to resort to Krylov methods only to invert the full~Jacobian. 

\subsection{Relevance for~Geophysics}

One might wonder about the applicability of this achievement to geophysical research. We begin by discussing the relevance of calculating traveling waves via Newton's method. Small Ekman numbers, like large Reynolds numbers, are associated with chaos and turbulence, not with the regularity and periodicity of traveling waves, which, moreover, are almost invariably unstable. 
The Boussinesq and Navier--Stokes equations generally contain a very large number of solution branches, most of which are partly or completely unstable, e.g.,   \cite{Schmiegel_PhD,gibson2009equilibrium,
pringle2009highly,boronska2010extreme_I,boronska2010extreme_II,zheng2024natural_1,zheng2024natural_2,zheng2025natural_3}. This profusion of states, sometimes called exact coherent structures, has become the focus of extensive research, motivated by the idea~\cite{Cvitanovic_Eckhardt,kawahara2012significance} that turbulence could be viewed from a dynamical systems perspective as a collection of trajectories ricocheting between periodic orbits along their unstable directions, a~kind of deterministic analogue of ergodic theory. 
Our Newton solver would enable the application of this line of research to geophysical flows. More generally, unstable states can cast light on the origin and organization of turbulent states; in one of the best known examples, ghostly Taylor vortices exist in even highly turbulent Taylor--Couette flow~\cite{lathrop1992transition,dong2007direct,huisman2014multiple,grossmann2016high} and the unstable underlying vortices may even reproduce the mean properties of the turbulent flow~\cite{eckhardt2020exact}.

We now turn to the relevance for geophysics of implicit Coriolis timestepping with large timesteps.
Constraints on timesteps (whether for stability or accuracy) arise from the different physical forces in the equations. 
In rapidly rotating fluids, the~Coriolis force generates inertial waves, which have a large range of frequencies and are continually generated and damped.
Using a timestep that greatly exceeds the constraint for resolving inertial waves can be considered analogous to using the incompressible Navier--Stokes equations (or the anelastic approximation), which act to filter out the high-frequency sound waves~\cite{van2024bridging,julien2024rescaled}. 

{ Using values of $\dt$ at which implicit Coriolis timestepping is possible but inaccurate can also help to explore the large parameter space. The~most economical procedure is to carry out a preliminary large-scale survey with a coarse spatial resolution and a large timestep. Interesting regions can then be accurately simulated using finer resolutions and smaller timesteps.} 

{ Previous research by two of the authors~\cite{feudel2011convection,feudel2013multistability,feudel2015bifurcations,feudel2017hysteresis} originated in the context of the Geoflow consortium, centered on microgravity experiments meant to mimic convective flows within the Earth~\cite{futterer2013sheet,Zaussinger_etal_2020}. These experiments were run in the Fluid Science Laboratory facility located in the European Space Agency's Columbus laboratory on the International Space Station. The~computations in~\cite{feudel2011convection,feudel2013multistability,feudel2015bifurcations,feudel2017hysteresis} used an explicit treatment of the Coriolis force and could not be continued to Ekman numbers smaller than $10^{-4}$. Geoflow has been succeeded by Atmoflow~\cite{TravnikovEgbers_2021}, whose purpose is to mimic convective flows in the atmosphere. We hope to use the new code to carry out computations relevant to AtmoFlow and to be able to achieve lower Ekman numbers.}

\section{Conclusions}
\label{sec:conclusion}

We have developed a pair of codes for simulating thermal convection in a rotating spherical fluid shell that relies on an implicit treatment of the Coriolis force. The~numerical cost of this improvement is quite manageable: the block diagonal matrix systems which arise from the implicit treatment of the diffusive terms must be replaced by block pentadiagonal matrix systems, which can still be solved by block banded LU decomposition and backsolving.
Once implemented in a timestepping code, implicit integration with a very large pseudo-timestep ($\dt=200$) can be leveraged to precondition the large linear systems that are the core of Newton's method. When only the diffusive terms are treated implicitly, this is known as Stokes preconditioning; the method developed here could be called Stokes-and-Coriolis preconditioning. 

We demonstrated this method's capabilities by carrying out continuation in Rayleigh number at various values of the Ekman number on the order of $10^{-5}$ for rotating waves with azimuthal wavenumbers 4, 8, and~12. We found several intriguing examples of branches containing plateaus in drift frequency, separated by intervals of rapid change delimited by pairs of saddle-node bifurcations. The~physical, or~at least phenomenological, reasons for these properties remain to be discovered.
We have also implemented continuation in the Ekman number, spaced logarithmically, in~which we automatically measure and increase the resolution as needed. We are unaware of any previous examples of continuation in the Ekman number or with automatic resolution adjustment in the~literature. 

We have measured the economy that is realized by the implicit treatment of the Coriolis force. For~Newton's method, the~advantage of the implicit over the explicit treatment is dramatic. For~example, the~explicit algorithm takes 20 times as many GMRES iterations (and hence 20 times as much CPU time) as the implicit method to compute states on the RW$_{12}$ branch at $\Ek=3\times 10^{-5}$. 
For lower Ekman numbers, the~explicit method takes an unmanageable number of GMRES iterations, making it impossible to use in~practice.

For timestepping, the~advantage of the implicit over the explicit treatment is equally spectacular. For~computing rotating waves at $\Ek=10^{-5}$, the~implicit algorithm can reasonably use timesteps that are almost 20 times past the temporal stability limit, meaning that simulations in this regime can be 20 times faster than if an explicit method were used.
In conclusion, the~implicit treatment of the Coriolis force greatly improves the efficiency of computations in the low Ekman number regime.


{\bf Code availability:} The code is in active development and fully available under the GNU General Public License v2.0 on the Github repository \url{https://github.com/JuanSembla/spherical_code_pmmh}.


{\bf Funding:} The calculations for this work were performed using high performance computing resources provided by the Grand Equipement National de Calcul Intensif (GENCI) at the Institut du D\'eveloppement et des Ressources en Informatique Scientifique (IDRIS, CNRS) through Grant No. A0162A01119.

{\bf Acknowledgments:} We gratefully acknowledge Alan Riquier for programming assistance and we thank Adrian van Kan, Benjamin Favier, and Rainer Hollerbach for helpful discussions.

\appendix
\section[\appendixname~\thesection]{}
Here, we present the intermediate calculations leading from \eqref{eq:impcor_e_init} and  \eqref{eq:impcor_f_init} to \eqref{eq:impcor_e_final} and \eqref{eq:impcor_f_final}. 
Using \eqref{eq:Ylm_norm}, the~recursion relations \eqref{eq:recu_all} for the associated Legendre polynomials can be rewritten in terms of the spherical harmonics~as
\begin{subequations}
\label{eq:recu_all_b}
\begin{align}
    \sin{\theta} \frac{d}{d\theta}\: Y_l^{|m|} &= \frac{l(l-|m|+1)}{2l+1} \;\frac{N_l^m}{N_{l+1}^m}\;Y_{l+1}^{|m|} - \frac{(l+|m|)(l+1)}{2l+1}\;\frac{N_l^m}{N_{l-1}^m}\;Y_{l-1}^{|m|}
    \label{eq:recu1_b}\\
    \cos\theta\; Y_l^{|m|} &= \frac{l-|m|+1}{2l+1}\;\frac{N_l^m}{N_{l+1}^m}\;Y_{l+1}^{|m|} + \frac{(l+|m|)}{2l+1}\;\frac{N_l^m}{N_{l-1}^m}\;Y_{l-1}^{|m|}
    \label{eq:recu2_b}
\end{align}
\end{subequations}
Applying \eqref{eq:recu_all_b} to \eqref{eq:impcor_e_init} leads to 
\begin{align*}
         \boldsymbol{e_{r}} \cdot &\curl \bF^{\rm implicit} \hspace*{9cm}\\
   =\sum_{m,l} &\left [ \frac{l(l+1)}{r^2} \Ek \left( \frac{\pd}{\pd t} - L_{l} \right) -\frac{2im}{r^2} \right] e_{lm} Y_l^m \\
     &+ \frac{2}{r^2} \left( \frac{l(l+1)}{r}-\frac{\pd}{\pd r} \right) f_{lm}  \left[\frac{l(l-|m|+1)}{2l+1} \;\frac{N_l^m}{N_{l+1}^m}\;Y_{l+1}^{|m|}
     - \frac{(l+|m|)(l+1)}{2l+1}\;\frac{N_l^m}{N_{l-1}^m}\;Y_{l-1}^{|m|} \right]\\
     & + 2\frac{l(l+1)}{r^2} \left( \frac{2}{r}-\frac{\pd}{\pd r} \right) f_{lm} \left[\frac{l-|m|+1}{2l+1}\;\frac{N_l^m}{N_{l+1}^m}\;Y_{l+1}^{|m|}
     + \frac{(l+|m|)}{2l+1}\;\frac{N_l^m}{N_{l-1}^m}\;Y_{l-1}^{|m|}\right]
     \end{align*}
Grouping terms in $Y_{l+1}^{|m|}$ and $Y_{l-1}^{|m|}$, we obtain
\begin{align*}
  = \sum_{m,l} &\left [ \frac{l(l+1)}{r^2} \Ek \left( \frac{\pd}{\pd t} - L_{l} \right) -\frac{2im}{r^2} \right] e_{lm}Y_l^m\\
 &   + \frac{2}{r^2} \frac{l(l-|m|+1)}{2l+1}  \left [ \left( \frac{l(l+1)}{r}-\frac{\pd}{\pd r} \right) + (l+1) \left( \frac{2}{r}-\frac{\pd}{\pd r} \right)  \right]\frac{N_l^m}{N_{l+1}^m} f_{lm} Y_{l+1}^m
     \\
  &   + \frac{2}{r^2} \frac{(l+|m|)(l+1)}{2l+1} \left [- \left( \frac{l(l+1)}{r}-\frac{\pd}{\pd r} \right) + l \left( \frac{2}{r}-\frac{\pd}{\pd r} \right) \right] \frac{N_l^m}{N_{l-1}^m}f_{lm} Y_{l-1}^m\\\\
     = \sum_{m,l}& \left [ \frac{l(l+1)}{r^2} \Ek\left( \frac{\pd}{\pd t} - L_{l} \right) -\frac{2im}{r^2} \right] e_{lm}Y_l^m\\
 &   + \frac{2}{r^2} \frac{(l-|m|+1)l(l+2)}{2l+1}  \left [ \frac{(l+1)}{r}-\frac{\pd}{\pd r}   \right]\frac{N_l^m}{N_{l+1}^m} f_{lm} Y_{l+1}^m
     \\
  &   + \frac{2}{r^2} \frac{(l+|m|)(l+1)(l-1)}{2l+1} \left [-\frac{l}{r}-\frac{\pd}{\pd r} \right] \frac{N_l^m}{N_{l-1}^m}f_{lm} Y_{l-1}^m
\end{align*}
Finally, changing the names of the indices so that all terms correspond to the same spherical harmonic $Y_l^m$, we obtain Equation \eqref{eq:impcor_e_final}:
\begin{align}
  =\sum_{m,l} &Y_l^m \left\{\left [ \frac{l(l+1)}{r^2} \Ek \left( \frac{\pd}{\pd t} - L_{l} \right) -\frac{2im}{r^2} \right] e_{lm}\right.\nonumber\\
 &   + \frac{2}{r^2} \frac{(l-1)(l+1)(l-|m|)}{2l-1}  \left [  \frac{l}{r}-\frac{\pd}{\pd r}   \right]\frac{N_{l-1}^m}{N_l^m} f_{l-1,m}
    \nonumber \\
  &  \left. + \frac{2}{r^2} \frac{(l+1+|m|)l(l+2)}{2l+3} \left [ -\frac{(l+1)}{r}-\frac{\pd}{\pd r} \right] \frac{N_{l+1}^m}{N_l^m}f_{l+1,m}\right\}
\end{align}

The same calculations can be applied to \eqref{eq:impcor_f_init} to obtain \eqref{eq:impcor_f_final}:
     \begin{align}
       \boldsymbol{e_{r}} \cdot \curl &\curl \bF^{\rm implicit} \hspace*{9cm}\nonumber\\
=\sum_{m,l} &- \left [ \frac{l(l+1)}{r^2} \Ek\left(\frac{\pd}{\pd t} - L_{l} \right) -\frac{2im}{r^2} \right] L_{l} f_{\rm lm} Y_{l}^m  \nonumber\\
   &  + \frac{2}{r^2} \left( \frac{l(l+1)}{r}-\frac{\pd}{\pd r} \right) e_{lm} \left[ \frac{l(l-|m|+1)}{2l+1} \;\frac{N_l^m}{N_{l+1}^m}\;Y_{l+1}^{|m|}
   - \frac{(l+|m|)(l+1)}{2l+1}\;\frac{N_l^m}{N_{l-1}^m}\;Y_{l-1}^{|m|}\right] \nonumber\\
   &  + 2\frac{l(l+1)}{r^2} \left( \frac{2}{r}-\frac{\pd}{\pd r} \right) e_{lm} \left[\frac{l-|m|+1}{2l+1}\;\frac{N_l^m}{N_{l+1}^m}\;Y_{l+1}^{|m|}
   + \frac{(l+|m|)}{2l+1}\;\frac{N_l^m}{N_{l-1}^m}\;Y_{l-1}^{|m|}\right]\nonumber\\\nonumber\\
 =\sum_{m,l}& - \left [ \frac{l(l+1)}{r^2} \Ek \left( \frac{\pd}{\pd t} - \: L_{l} \right) -\frac{2im}{r^2} \right] L_{l}f_{lm} Y_l^m \nonumber\\
&     + \frac{2}{r^2} \frac{l(l-|m|+1)}{2l+1}  \left [ \left( \frac{l(l+1)}{r}-\frac{\pd}{\pd r} \right) + (l+1) \left( \frac{2}{r}-\frac{\pd}{\pd r} \right)  \right] \frac{N_l^m}{N_{l+1}^m} e_{lm} Y_{l+1}^m\nonumber\\
 &    + \frac{2}{r^2} \frac{(l+|m|)(l+1)}{2l+1} \left [- \left( \frac{l(l+1)}{r}-\frac{\pd}{\pd r} \right) + l \left( \frac{2}{r}-\frac{\pd}{\pd r} \right) \right]  \frac{N_l^m}{N_{l-1}^m} e_{lm}Y_{l-1}^m\nonumber\\\nonumber\\
 =\sum_{m,l}& - \left [ \frac{l(l+1)}{r^2} \Ek \left( \frac{\pd}{\pd t} - L_{l} \right) -\frac{2im}{r^2} \right] L_{l}f_{lm} Y_l^m \nonumber\\
&     + \frac{2}{r^2} \frac{l(l-|m|+1)(l+2)}{2l+1}  \left [ \frac{(l+1)}{r}-\frac{\pd}{\pd r}  \right] \frac{N_l^m}{N_{l+1}^m} e_{lm} Y_{l+1}^m
  \nonumber   \\
 &    + \frac{2}{r^2} \frac{(l+|m|)(l+1)(l-1)}{2l+1} \left [ -\frac{l}{r}-\frac{\pd}{\pd r}   \right]  \frac{N_l^m}{N_{l-1}^m} e_{lm}Y_{l-1}^m
 \nonumber\\\nonumber\\
 =\sum_{m,l} &Y_l^m\left\{ - \left [ \frac{l(l+1)}{r^2} \Ek\left( \frac{\pd}{\pd t} - L_l \right) -\frac{2im}{r^2} \right] L_lf_{lm}\right. \nonumber\\
&     + \frac{2}{r^2} \frac{(l-1)(l-|m|)(l+1)}{2l-1}  \left [ \frac{l}{r}-\frac{\pd}{\pd r}  \right] \frac{N_{l-1}^m}{N_l^m} e_{l-1,m}
  \nonumber   \\
 &   \left.+ \frac{2}{r^2} \frac{(l+1+|m|)(l+2)l}{2l+3} \left [ -\frac{(l+1)}{r}-\frac{\pd}{\pd r}  \right]  \frac{N_{l+1}^m}{N_l^m} e_{l+1,m}\right\}
\end{align}

\end{document}